\documentclass[pra,superscriptaddress,twocolumn,showpacs,preprintnumbers,nofootinbib,amsmath,amssymb,longbibliography]{revtex4-2}

\usepackage{multirow}
\usepackage{graphicx}% Include figure files
\usepackage{dcolumn}% Align table columns on decimal point
\usepackage{bm}% bold math
\usepackage{psfrag}
\usepackage{physics}
\usepackage{epsfig}
\usepackage{amsmath}
\usepackage{amssymb}
\usepackage{MnSymbol}
\usepackage{color}
\usepackage{bbold}
\usepackage{bbm}
\usepackage[FIGTOPCAP,raggedright,nooneline]{subfigure}
\usepackage{verbatim}
 \usepackage[applemac]{inputenc}
\usepackage{textcmds} 
\usepackage{tikz}
\usepackage{natbib}
\usepackage[colorlinks=true,
            linkcolor=magenta,
            urlcolor=blue,
            citecolor=blue]{hyperref}

\definecolor{darkgreen}{rgb}{0,0.5,0}
\definecolor{redred}{HTML}{D53E4F}
\newcommand{\danger}[1]{{\color{redred} #1}}
\newcommand{\greencolor}[1]{{\color{greengreen} #1}}
\newcommand{\mtext}[1]{{\color{mage} #1}}
\newcommand{\gtext}[1]{{\color{darkgreen} #1}}

\newcommand{\rla}{\danger{r}}
\newcommand{\gla}{\gtext{g}}
\newcommand{\cn}{\centering}

\usepackage{orcidlink}
\usepackage{array}
\newcolumntype{L}[1]{>{\raggedright\let\newline\\\arraybackslash\hspace{0pt}}m{#1}}
\newcolumntype{C}[1]{>{\centering\let\newline\\\arraybackslash\hspace{0pt}}m{#1}}
\newcolumntype{R}[1]{>{\raggedleft\let\newline\\\arraybackslash\hspace{0pt}}m{#1}}

\begin{document}

\title{Digital quantum simulation of a (1+1)D SU(2) lattice gauge theory with ion qudits}

\author{Giuseppe Calaj\'o\orcidlink{0000-0002-5749-2224}}
\affiliation{Istituto Nazionale di Fisica Nucleare (INFN), Sezione di Padova, I-35131 Padova, Italy}

\author{Giuseppe Magnifico\orcidlink{0000-0002-7280-445X}}
\affiliation{Dipartimento di Fisica e Astronomia ``G. Galilei'', via Marzolo 8, I-35131 Padova, Italy}
\affiliation{Dipartimento di Fisica, Universit\`a di Bari, I-70126 Bari, Italy.}
\affiliation{Istituto Nazionale di Fisica Nucleare (INFN), Sezione di Bari, I-70125 Bari, Italy}

\author{Claire Edmunds\orcidlink{0000-0002-2367-9074}}
\affiliation{Universit\"at Innsbruck, Institut f\"ur Experimentalphysik, Technikerstra\ss e 25a, Innsbruck, Austria}

\author{Martin Ringbauer\orcidlink{0000-0001-5055-6240}}
\affiliation{Universit\"at Innsbruck, Institut f\"ur Experimentalphysik, Technikerstra\ss e 25a, Innsbruck, Austria}

\author{Simone Montangero\orcidlink{0000-0002-8882-2169}}
\affiliation{Dipartimento di Fisica e Astronomia ``G. Galilei'', via Marzolo 8, I-35131 Padova, Italy}
\affiliation{Padua Quantum Technologies Research Center, Universit\'a degli Studi di Padova}
\affiliation{Istituto Nazionale di Fisica Nucleare (INFN), Sezione di Padova, I-35131 Padova, Italy}

\author{Pietro Silvi\orcidlink{0000-0001-5279-7064}}
\affiliation{Dipartimento di Fisica e Astronomia ``G. Galilei'', via Marzolo 8, I-35131 Padova, Italy}
\affiliation{Padua Quantum Technologies Research Center, Universit\'a degli Studi di Padova}
\affiliation{Istituto Nazionale di Fisica Nucleare (INFN), Sezione di Padova, I-35131 Padova, Italy}

\date{\today}
 
\begin{abstract}
%The simulation of the dynamics of non-abelian lattice gauge theories (LGTs) at finite densities is a long standing goal  of classical and quantum computation. This ambitious goal is compounded by the necessity to maintain exact gauge invariance, often demanding complex computational schemes and high control over quantum hardware.
We present a quantum simulation strategy for a (1+1)D SU(2) non-abelian lattice gauge theory with dynamical matter, a hardcore-gluon Hamiltonian Yang-Mills, tailored to a six-level trapped-ion qudit quantum processor, as recently experimentally realized~\cite{ringbauer2022universal}.
We employ a  qudit encoding fulfilling gauge invariance, an SU(2) Gauss' law.  We discuss the experimental feasibility of generalized M{\o}lmer-S{\o}rensen gates used to efficiently simulate the dynamics. 
%We consider, as dynamical processing resources, generalized M{\o}lmer-S{\o}rensen gates, simultaneously addressing multiple qudit transitions on each involved ion, while discussing their experimental feasibility.
%Through a convenient qudit encoding which preserves that gauge invariance, 
%on the qudit
%and the implementation of generalized M{\o}lmer-S{\o}rensen gates, we
We illustrate how a shallow circuit with these resources is sufficient to implement scalable digital quantum simulation of the model. We also numerically show that this model, albeit simple, can dynamically manifest physically-relevant properties specific to non-abelian field theories, such as
%meson {\it and}
 baryon excitations.
\\%Finally we discuss the details the experimental feasibility of the proposal.\\
\end{abstract}

\maketitle

%\section{Introduction}
Gauge theories are a fundamental theoretical framework in physics playing an important role in many active areas of research spanning from high energies~\cite{kronfeld2012twenty} to condensed matter~\cite{fradkin2013field} and quantum information science~\cite{zeng2019quantum}.
Their formulation on a lattice, known as lattice gauge theories, (LGT)~\cite{wilson1974confinement,kogut1975hamiltonian,banks1976strong}, is particularly suited to study non-perturbative effects.
Monte Carlo techniques have been extremely successful over the years~\cite{gattringer2009quantum} in tackling various %non-perturbative  
models, including quantum chromodynamics, but they are limited by the sign problem to certain physical regimes, and struggle to capture
%the physics at finite matter density, especially
the real-time evolution outside equilibrium~\cite{calzetta2009nonequilibrium}.
In the last years, following the advances in quantum and quantum-inspired computation, alternative new possibilities to face these problems have emerged based either on tensor network numerical techniques~\cite{silvi2014lattice,dalmonte2016lattice,silvi2019tensor} or on analog and digital quantum simulation~\cite{wiese2013ultracold,banuls2020simulating, dimeglio2023quantum,halimeh2023cold}, involving different experimental platforms such as cold atoms ~\cite{banerjee2012atomic,zohar2015quantum,PhysRevA.95.023604,kuhn2014quantum,tagliacozzo2013optical,tagliacozzo2013simulation,yang2020observation,halimeh2305spin,aidelsburger2022cold}, superconducting circuits~\cite{marcos2013superconducting,mezzacapo2015non,atas20212,farrell2024quantum} and trapped ions~\cite{hauke2013quantum,martinez2016real,muschik2017u,nguyen2022digital,davoudi2020towards}. While with the numerical approach both abelian~\cite{rico2014tensor,pichler2016real, PhysRevD.98.074503,magnifico2020real,zhang2023observation,mildenberger2022probing,rigobello2021entanglement, PhysRevD.107.014505, PhysRevD.108.014504, PhysRevD.108.014516, belyansky2023highenergy, kebric2023confinement, florio2023mass} and non-abelian~\cite{kuhn2015non,silvi2017finite,banuls2017efficient, cataldi202321d, rigobello2023hadrons, hayata2023dense} lattice gauge theory models have been theoretically studied, early demonstrations of quantum simulation of LGT largely focused on abelian theories~\cite{martinez2016real,nguyen2022digital,zhou2022thermalization,paulson2021simulating, angelides2023firstorder, chai2023entanglement, davoudi2024scattering,farrell2023scalable}.
First attempts to tackle non-abelian models have been proposed relying ether on hybrid quantum computation schemes such as variational eigensolvers~\cite{atas20212}, % or concentrating on the implementation of {\color{red} small lattice sizes} preserving the gauge invariance
while other quantum simulation encoding proposals are fairly limited in system size for realistic experimental platforms based on qubits~\cite{ciavarella2021trailhead,farrell2023preparations,PhysRevD.107.054512, PhysRevResearch.5.033184, PhysRevD.108.094513,ciavarella2024quantum}.
\begin{figure}
\centering
\includegraphics[width=0.5\textwidth]{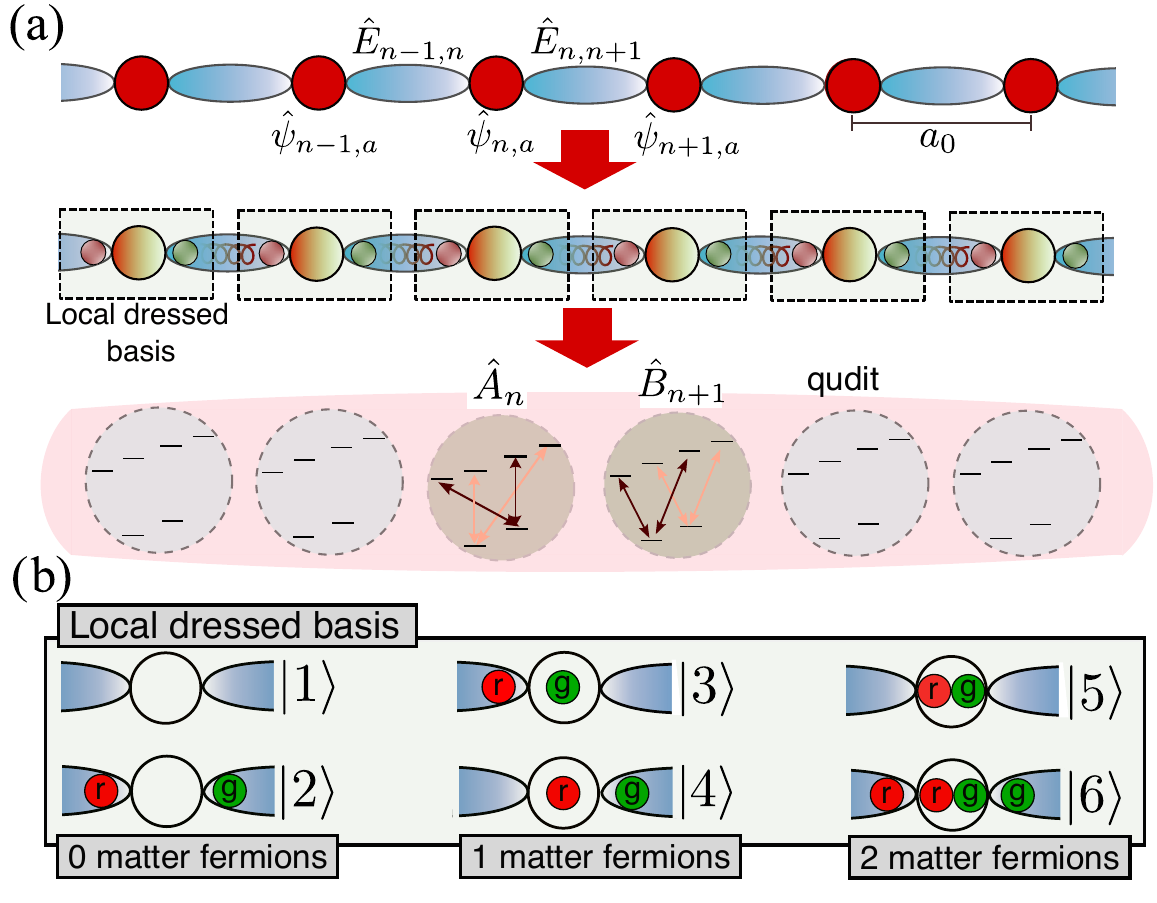}%
\caption{\textit{Sketch of the model.}--(a)  A (1+1)D Yang-Mills SU(2) lattice gauge model is encoded in a linear chain of trapped ion qudits. The lattice gauge links (blue ovals) connecting the matter sites (red circles) are truncated to a 5-dimensional Hilbert space. The mapping is then performed by exploiting the local dressed basis~\eqref{eq:dressed basis}, obtained through the decomposition of the gauge links into a pair of rishons. 
(b) Pictorial representation of the local dressed basis given in Eq.~\eqref{eq:dressed basis}. The red and green circles in the links represent the two colored rishons while the ones in the sites represent the two colored matter fermions.}
\label{fig:setup}
\end{figure}

One potential avenue to perform large-scale quantum computation relies on qudit-based quantum processors, which have been proposed in several platforms such as Rydberg arrays~\cite{kruckenhauser2022high,cohen2021quantum}, photonic circuits \cite{chi2022programmable} and ultra-cold atomic mixture~\cite{kasper2021universal}.
An experimental breakthrough along this direction has been the demonstration of a universal 7-level optical qudit quantum processor implemented on a chain of trapped $^{40}\textrm{Ca}^+$ ions~\cite{ringbauer2022universal}. Using other ion species, the qudit dimension can be further extended as suggested by recent results with up to 13 levels~\cite{low2023control}.
These hardware developments have already stimulated interesting proposals~\cite{gonzalez2022hardware,zache2023fermion,popov2023variational} and experiments~\cite{methSimulating2DLattice2023,zalivako2024towards} for performing simulations of lattice gauge theories exploring this enlarged Hilbert space. 
However, a proof-of-principle experimental demonstration of a scalable quantum simulation of the dynamics of non-abelian LGTs is still lacking.  Such demand is of fundamental importance not only to build new intuitions  on not fully understood mechanisms of high-energy physics~\cite{kronfeld2012twenty} but also to explore phenomena in condensed matter models exhibiting non-abelian topological order~\cite{xu2012topological,kalinowski2023non}.

Here, we present a compact rishon representation for a truncated Yang-Mills SU(2) (1+1)D lattice gauge theory  reduced to a 6-dimensional local Hilbert space, embedding the gauge and fermionic degrees of freedom.
This representation allows us to preserve the non-abelian gauge symmetry,  remove the fermionic character of matter (de-fermionization), and maintain the range of the gauge-matter interaction to nearest neighbors. This is in contrast with other approaches that rely either on the elimination of the gauge~\cite{muschik2017u} or the matter~\cite{popov2023variational} degree of freedom, which are not straightforwardly extendable to treat non-abelian models, or on a fermionic-qudit quantum processor~\cite{gonzalez2023fermionic,zache2023fermion} whose proposal is currently limited to programmable Rydberg array platforms.
%while maintaining the range of the gauge-matter interaction to the nearest neighbors, in contrast with approaches where the adiabatic elimination of the gauge degree of freedom requires long-range interaction terms~\cite{muschik2017u}.
In this work, we present an experimentally feasible proposal based on a currently available $^{40}$Ca$^+$ trapped-ion qudit quantum processor~\cite{ringbauer2022universal,methSimulating2DLattice2023}. In particular, we demonstrate how a digital quantum simulation of the model can be efficiently implemented by making use of generalized M{\o}lmer-S{\o}rensen gates (MS) realized by simultaneous driving of multiple transitions~\cite{low2020practical}. In this manner, we obtain a shallow circuit for each time step enabling a feasible digital quantum simulation on current devices capable of distinguishing between bare {\it meson} and {\it baryon} production and thus signaling the non-abelian nature of the model. 
Finally, we discuss the major challenges for the experimental realization of the proposal, but we demonstrate that its implementation is compatible with current technologies.

This paper is structured as follows. In Sec.~\ref{Sec.Model}, we introduce the (1+1)D SU(2) Yang-Mills lattice gauge model and derive the truncated qudit Hamiltonian defined on a local six-dimensional Hilbert space. In Sec.~\ref{Sec:encoding}, we present the strategy used to encode the model on a trapped-ion qudits quantum processor and the generalized M{\o}lmer-S{\o}rensen gates used in the quantum digital simulation. In Sec.~\ref{Sec.phen}, we show a few paradigmatic examples of the non-abelian dynamics inherent in the model. In Sec.~\ref{Sec.Dig_sim}, we present the strategy used to perform a quantum digital simulation of the model dynamics. In Sec.~\ref{Sec.experiment}, we discuss the experimental feasibility of the proposal. Finally, in Sec.~\ref{sec.conclusion}, we summarize our results and discuss future directions of research.

\section{SU(2) Lattice Yang-Mills}\label{Sec.Model}

We start by considering a Hamiltonian Yang-Mills lattice gauge model, with SU(2) color symmetry, in one spatial dimension, focusing on the low energy regime. The model illustrates a (flavorless) fermionic matter field coupled to an SU(2) gauge field, as depicted in Fig.~\ref{fig:setup}(a).
The matter field, representing quarks of bare mass $m_0$, is described by staggered fermions $\hat\psi_{na}$ \cite{susskind1977lattice}, with two colors $a \in\{\rla, \gla\}$ (say red and green), living on the lattice sites $n$, and satisfying standard Dirac anticommutation
rules $\{\hat\psi_{na},\hat\psi^{\dagger}_{n'b}\}=\delta_{n,n'} \delta_{a,b}$. Conversely, the non-abelian gauge field lives on the lattice bonds between sites $n$ and $n+1$. %(its algebra $\hat U^{ab}$, $\hat E^2$ is defined later on).
Following the Kogut-Susskind formulation of gauge fields on a lattice~\cite{kogut1975hamiltonian}, the system Hamiltonian reads:
\begin{equation}\label{eq:H}
\begin{split}
\hat H_{0}=&\frac{c\hbar}{2a_0}\sum_n\sum_{a,b=\rla, \gla}\left[-i\hat\psi^{\dagger}_{na}\hat U^{ab}_{n,n+1}\hat\psi_{n+1b}+{\rm H.c.} \right]\\&+m_0c^2\sum_{na}(-1)^{n} \hat\psi^{\dagger}_{na}\hat\psi_{na}+ g_0^2 \frac{c\hbar}{2a_0}\sum_n\hat E^2_{n,n+1}
\end{split}
\end{equation}
where $a_0$ is the lattice spacing, $\hbar$ the Planck constant and $c$ the relativistic speed of free massless particles.
The first two terms of Eq.~\eqref{eq:H} describe the lattice Hamiltonian of the covariant Dirac equation for massive quarks. It uses the staggered mass term to create two sublattices, representing each a component of the two-spinor Dirac field,  avoiding the doubling problem \cite{susskind1977lattice}. In this sense, quarks are staggered fermions on even sites, while anti-quarks are staggered fermion holes on odd sites.

The last term is the pure gauge Hamiltonian, which contains only an electric component $\hat E^2_{n,n+1}$ since there are no magnetic fields in one spatial dimension. The dimensionless coupling $g^2_0$ depends on the quark color charge $q_c$ and on the lattice spacing $a_0$: In one spatial dimension, it  scales as
%$g^2_0 = \frac{q_c^2 a_0^2}{c \hbar \epsilon_c}$
$g^2_0 = ({q_c^2 a_0^2})/({c \hbar \epsilon_c})$
with $\epsilon_c$ the vacuum color permittivity,
assuming the theory is super-renormalizable.
The SU(2)-electric field energy density is captured by the (dimensionless) quadratic Casimir operator
%The first  describes fermion-hopping between nearest-neighboring sites mediated by the gauge field, living on the lattice links, via the SU(2)-parallel transporter operator $\hat U^{ab}_{j,j+1}$. The second term takes into account the fermion masses and the staggered sign ensures to avoid fermion doubling \cite{susskind1977lattice}. Finally, the last contribution represents the SU(2)-electric energy density given by the quadratic Casimir operator on every link 
$\hat E^2_{n,n+1}=| \hat{\mathbf{L}}_{n,n+1}|^2=|\hat{\mathbf{R}}_{n,n+1}|^2$,
%Note that this coupling depends on the lattice spacing (linearly in 1D).
where the algebra operators $\hat L^{(\nu)}_{n,n+1}$ and $\hat R^{(\nu)}_{n,n+1}$ with coordinates $\nu\in\{x,y,z\}$ are respectively the left and right group generators of the gauge transformation on each link. The gauge-field algebra is defined by the commutation rules
\begin{equation} 
\begin{aligned}
\left[ \hat{L}^{(\nu)}_{n,n+1},\hat{R}^{(\nu')}_{n',n'+1} \right]&=0  \\
\left[\hat{L}^{(\nu)}_{n,n+1},\hat{L}^{(\nu')}_{n',n'+1} \right]&=i\delta_{nn'}\epsilon^{\nu\nu'\nu''}\hat{L}^{(\nu'')}_{n,n+1}\\
\left[\hat{R}^{(\nu)}_{n,n+1},\hat{R}^{(\nu')}_{n',n'+1} \right]&=i\delta_{nn'}\epsilon^{\nu\nu'\nu''}\hat{R}^{(\nu'')}_{n,n+1}\\
\left[\hat{L}^{(\nu)}_{n,n+1},\hat U^{ab}_{n',n'+1} \right]&=-\delta_{nn'}\sum_c\frac{\sigma^{(\nu)}_{ac}}{2}\hat U^{cb}_{n,n+1}\\
\left[\hat{R}^{(\nu)}_{n,n+1},\hat U^{ab}_{n',n'+1} \right]&=\delta_{nn'}\sum_c\hat U^{ac}_{n,n+1}\frac{\sigma^{(\nu)}_{cb}}{2}
\end{aligned}
\end{equation}
where $\epsilon^{\nu\nu'\nu''}$ is the Levi-Civita symbol for SU(2) and $\sigma^{(\nu)}$ are the Pauli matrices.

\begin{table*}[t]
 \begin{tabular}{|p{110pt}|p{132pt}|p{132pt}|p{110pt}|}
 \hline
  \cn $\hat A^{(1)}$ & \cn $\hat A^{(2)}$ & \cn $\hat B^{(1)}$ &
  $\qquad \qquad \quad \hat B^{(2)}$
  \\
  \hline
  \cn $
  \begin{pmatrix}
   0      &  &  &\sqrt{2} & & \\
         & 0  & 1 & & & \\
    & 1  & 0 & & & 1\\
   \sqrt{2}  &   &  & 0 & \sqrt{2} & \\
     &   &  &  \sqrt{2} & 0 & \\
     &   & 1 & & & 0\\
\end{pmatrix}$
&
\cn $
\begin{pmatrix}
   0      &  &  &\sqrt{2}i & & \\
         & 0  & i & & & \\
    & -i  & 0 & & & i\\
   -\sqrt{2}i  &   &  & 0 & \sqrt{2}i & \\
     &   &  &  -\sqrt{2}i & 0 & \\
     &   & -i & & & 0\\
\end{pmatrix}$
&
 \cn $
\begin{pmatrix}
   0      &  &  -\sqrt{2}i & & & \\
         & 0  &  & -i &  & \\
   \sqrt{2}i  &   & 0 & & -\sqrt{2}i  & \\
  &  i &  & 0 &  & -i \\
     &   & \sqrt{2}i  &  & 0 & \\
     &   &  & i & & 0\\
\end{pmatrix}$
&
 $\quad
\begin{pmatrix}
   0      &  &  \sqrt{2} & & & \\
         & 0  &  & 1 &  & \\
   \sqrt{2}  &   & 0 & & \sqrt{2} & \\
  &  1 &  & 0 &  & 1\\
     &   & \sqrt{2}  &  & 0 & \\
     &   &  & 1 & & 0\\
\end{pmatrix}$
\\ \hline
\end{tabular}
\caption{ \label{tab:effmatrices1}
 Four of the six matrices that define the effective model on qudits. These four are related to the covariant Dirac transport, and they are factors of a nearest-neighbor interaction $\hat A^{(1)}_n \hat B^{(1)}_{n+1} + \hat A^{(2)}_n \hat B^{(2)}_{n+1}$.
}
\end{table*}

\begin{table*}[t]
 \begin{tabular}{|p{110pt}|p{110pt}|p{132pt}|p{132pt}|}
  \hline
  \cn $\hat M$ & \cn $\hat C$ & \cn $\hat D^{(L)}$ & $\qquad \qquad \qquad \hat D^{(R)}$ \\
  \hline
  \cn $
  \begin{pmatrix}
   0      &  &  & & & \\
         & 0  &  & & & \\
    &   & 1 & & & \\
 &   &  & 1 &  & \\
     &   &  &  & 2 & \\
     &   &  & & & 2\\
\end{pmatrix}$
&
  \cn $
  \begin{pmatrix}
   0      &  &  & & & \\
         & 2  &  & & & \\
    &   & 1 & & & \\
 &   &  & 1 &  & \\
     &   &  &  & 0 & \\
     &   &  & & & 2\\
\end{pmatrix}$
&
  \cn $
  \begin{pmatrix}
   +1      &  &  & & & \\
         & -1  &  & & & \\
    &   & +1 & & & \\
 &   &  & -1 &  & \\
     &   &  &  & +1 & \\
     &   &  & & & -1\\
\end{pmatrix}$
&
  $\quad \; 
  \begin{pmatrix}
   +1      &  &  & & & \\
         & -1  &  & & & \\
    &   & -1 & & & \\
 &   &  & +1 &  & \\
     &   &  &  & +1 & \\
     &   &  & & & -1\\
\end{pmatrix}$
\\ \hline
 \end{tabular}
\caption{ \label{tab:effmatrices2}
 The last two matrices that define the effective model, related to mass term (before staggerization) and chromoelectric energy density respectively. The table also shows the matrices of the link fermion parity selection rule.
}
\end{table*}

To represent these operators in a matrix form it is useful to express them in the chromoelectric basis of states $|j m_L m_R \rangle$, where $j \in \mathbb{N}/2$ labels the spin irreducible representations (irreps), $m_R \in \{-j,..,j\}$ labels a spin shell $j$, and $m_L \in \{-j,..,j\}$ labels a state within the spin shell adjoint to $j$.% (identical to $j$ for SU(2)).
The gauge field algebra operators in this basis~\cite{ZoharBurrelloPRD}
\begin{equation} 
\begin{aligned}
 \langle j' m'_L m'_R |
  \hat E^2
 | j m_L m_R \rangle &=
 j(j+1)
 \delta_{j,j'} \delta_{m_L,m_L'} \delta_{m_R,m_R'}
 \\
 \langle j' m'_L m'_R |
  \hat U^{ab}
 | j m_L m_R \rangle &=
 \left(C^{j,m_L}_{j',m'_L;\frac{1}{2},a}\right)^*
 C^{j',m'_R}_{j,m_R;\frac{1}{2},b}
\end{aligned}
\end{equation}
where $C^{J,M}_{j_1,m_1;j_2,m_2} = \langle j_1,m_1;j_2,m_2 | J, M \rangle$ are the Clebsh-Gordan coefficients, i.e.~the fusion rules, for SU(2).

The model \eqref{eq:H} is designed to be symmetry-invariant
under the gauge transformations generated by
$\hat{\mathbf{G}}_{n} = \left(\hat{\mathbf{R}}_{n-1,n} + \hat{\mathbf{S}}_{n} + \hat{\mathbf{L}}_{n,n+1}\right)$, where
$\hat{S}^{(\nu)}_{n} = \frac{1}{2} \sum_{a,b} \sigma^{(\nu)}_{ab}\hat\psi^{\dagger}_{na}\hat\psi_{nb}$ generates the color-rotations for the quarks. Under this observation, the non-abelian Gauss' law, which defines the sector of physical states $|\Psi_{\text{phys}}\rangle$, reads
\begin{equation} 
 \left| \hat{\mathbf{G}}_{n} \right|^2 |\Psi_{\text{phys}} \rangle = 0 \;\;\forall n
 \qquad \mbox{(Gauss' Law)}
 \label{eq:gauss}
\end{equation}
corresponding to the absence of a color background.

\subsection{Hardcore gluon approximation}\label{Sec.ModelB}

To  digitally quantum simulate Hamiltonian \eqref{eq:H} it is necessary to truncate the infinite local gauge Hilbert space to a finite dimension. This can be done at low energies by employing the Quantum Link Model (QLM)~\cite{chandrasekharan1997quantum} formalism,
where only a finite set of shells $j$ are considered.
Basically, a cutoff is chosen for the $\hat E^2$ energy term, and the spin shells $j$ above the cutoff are discarded. Our proposal for digital quantum simulation considers a QLM cutoff that includes the $j=0$ and $j=\frac{1}{2}$ shells, that is, the smallest cutoff which allows quarks to form bound states of baryons and mesons. In practice in our descriptions we keep all the states that are reachable from the bare vacuum with a single application of $\hat U^{ab}$: in analogy to cold atom physics, we refer to this cutoff strategy as ``hardcore gluons'', %\qq{hardcore gluons}.
and it is a reasonable approximation for strong coupling regimes~\cite{cataldi202321d,rigobello2023hadrons,Yannick_fades_away}.
Ultimately, this picture describes a 5-dimensional gauge field Hilbert space at each bond, spanned by the link basis set $\{|00\rangle,|\rla\rla\rangle,|\gla\gla\rangle,|\gla\rla\rangle,|\rla\gla\rangle\}$. Within this representation it is possible to decompose each gauge field bond into a pair of exotic fermion sub-orbitals (rishons). Then, the parallel transporter can be efficiently decomposed as
%the parallel transporter operator in terms of a pair of rishon operators
$\hat U^{ab}_{n,n+1}=\frac{1}{\sqrt{2}}(\hat \zeta^a_{n,n+1})_L(\hat \zeta^{b\dagger}_{n,n+1})_R$  where each of the exotic fermions $\hat \zeta^a$
lives on the left (L) and right (R) rishon of the same link, as sketched in the second row of Fig.~\ref{fig:setup}(a).
Each rishon mode is ultimately 3-dimensional, spanned by the state $|0\rangle = |j=0,m=0\rangle$ with even fermion parity $\langle \hat{P} \rangle =+1$, and the states $|\rla\rangle = |j=\frac{1}{2},m=+\frac{1}{2}\rangle,$ $|\gla\rangle = |j=\frac{1}{2},m=-\frac{1}{2}\rangle$ with odd fermion parity $\langle \hat{P} \rangle=-1$.
In this representation, the fermion parity on a link becomes an abelian (gauge $\mathbb{Z}_2$) symmetry of the dynamics, which  defines the 5-dimensional space.
%and by requesting it to be even on each link
%\begin{equation}
% \left[ (\hat{P}_{n,n+1})_{L} (\hat{P}_{n,n+1})_{R} - 1 \right]
 %|\Psi_{\text{phys}} \rangle = 0
 %\;\; \forall n \quad \mbox{(Link Law)}
%\end{equation} 
%we restore the 5-dimensional space as above.
By contrast, the matter sites are regular Dirac fermions, thus spanned by the site basis: $|0\rangle$ the fermion vacuum (even fermion parity), $|\rla\rangle = \hat \psi^{\dagger}_{\rla} |0\rangle$ and $|\gla\rangle = \hat \psi^{\dagger}_{\gla} |0\rangle$ singly-occupied states (odd parity), and finally
$|d\rangle = \hat \psi^{\dagger}_{\rla} \hat \psi^{\dagger}_{\gla} |0\rangle$ the doubly-occupied state (even). We can now fuse together the (R)$_{n-1,n}$ rishon, the matter at site $n$, and the (L)$_{n,n+1}$ rishon into a unique dressed site. We now enforce the non-Abelian Gauss' Law of Eq.~\eqref{eq:gauss}, resulting in an effective 6-dimensional gauge-invariant basis for the dressed site defined 
as the tensor product of the  matter field on a lattice site and of the rishon states on its left and right. This basis, pictorially sketched in Fig.~\ref{fig:setup}(b),  contains only states with even total fermion parity  and explicitly reads
%and taking values in the basis $\{|0\rangle,|\rla\rangle,|\gla\rangle$. As described in details in App.~\ref{App_A}, we employ this representation to dress every physical matter site, spanned by the site basis set $\{|0\rangle,|\rla\rangle,|\gla\rangle,|\gla\rla\rangle$, with the rishons living in the adjacent gauge links. By constraining the link dynamics to respect the  SU(2) nonabelian gauge symmetry of the model, we are left with the following dressed basis of dimension $6$, pictorially represented in Fig.~\ref{fig:setup}(b),
\begin{equation}\label{eq:dressed basis}
\begin{aligned}
 | 1\rangle &=|0,0,0\rangle 
 & | 2\rangle&=\frac{|\rla,0,\gla\rangle-|\gla,0,\rla\rangle}{\sqrt{2}} \\
 | 3\rangle &=\frac{|\gla,\rla,0\rangle-|\rla,\gla,0\rangle}{\sqrt{2}}
 &| 4\rangle &=\frac{|0,\rla,\gla\rangle-|0,\gla,\rla\rangle}{\sqrt{2}} \\
 | 5\rangle &=|0,d,0\rangle
 & |6\rangle &=\frac{|\rla,d,\gla\rangle-|\gla,d,\rla\rangle}{\sqrt{2}} \,.
\end{aligned}
\end{equation}
 As such, the model preserves fermion parity at each dressed site~\cite{Erez_eliminates_fermions}. % which leads to elimination of the fermionic matter .
In this dressed site basis, it is relatively straightforward  (see App.~\ref{App_A}) to rewrite the hardcore-gluon SU(2) Yang-Mills model from~\eqref{eq:H} into an effective Hamiltonian
\begin{equation}\label{eq:Heff}
\hat H= \sum_n\left[\hat A^{(1)}_n\hat B^{(1)}_{n+1}+\hat A^{(2)}_n\hat B^{(2)}_{n+1}\right]
+m\sum_n(-1)^{n}\hat M_n+g^2\sum_n\hat C_n
\end{equation}
where we set the energy scale based on the hopping term, by rescaling $\hat H_{0} \to \hat H = \frac{4 \sqrt{2} a_0}{c \hbar} \hat H_{0}$, to work in natural units, with a dimensionless Hamiltonian $\hat H$ and dimensionless couplings
%$J=\frac{\sqrt{2}}{8}$,
$m = \frac{4 \sqrt{2} m_0 a_0 c}{\hbar}$, and
$g^2 = \frac{3 \sqrt{2}}{4} g_0^2$. The $6 \times 6$ matrix operators appearing in this expression are reported in Tables~\ref{tab:effmatrices1} and~\ref{tab:effmatrices2}.

While the non-Abelian Gauss' law has been already enforced in the definition of a dressed basis, we still have to take into account the Link Law which arose from splitting a gauge field into two rishons. In the effective model the link law translates into an abelian selection rule
\begin{equation}
 \left( \hat{D}_{n}^{(L)} \hat{D}_{n+1}^{(R)} - 1 \right)
 |\Psi_{\text{phys}} \rangle = 0
 \;\; \forall n \quad \mbox{(Link Law)}.
\end{equation} 
This symmetry at each link is protected by the Hamiltonian, thus in principle, it would be sufficient to satisfy the constraint on the initial state of the dynamics. However, this symmetry may be disrupted by noise and imperfections in a digital quantum simulator~\cite{martinez2016real,nguyen2022digital}. In Sec.~\ref{Sec.post} we discuss how to mitigate this error via a postselection procedure. %thus it becomes important to preserve it by the algorithm %(see Sec.~\ref{Sec.post} later on).

Another important symmetry to discuss is the conservation of the total baryon number $\hat{N_b} = \frac{1}{2} \sum_n (\hat{M}_n - 1)$. This quantity can be controlled by appropriately constructing the starting state of the dynamics and allows the quantum simulator to explore areas of the phase diagram, with high baryon density, inaccessible to Monte Carlo simulations due to the  sign problem.

\section{Encoding into  trapped-ion qudits}\label{Sec:encoding} 
\begin{figure}
    \includegraphics[width=0.45\textwidth]{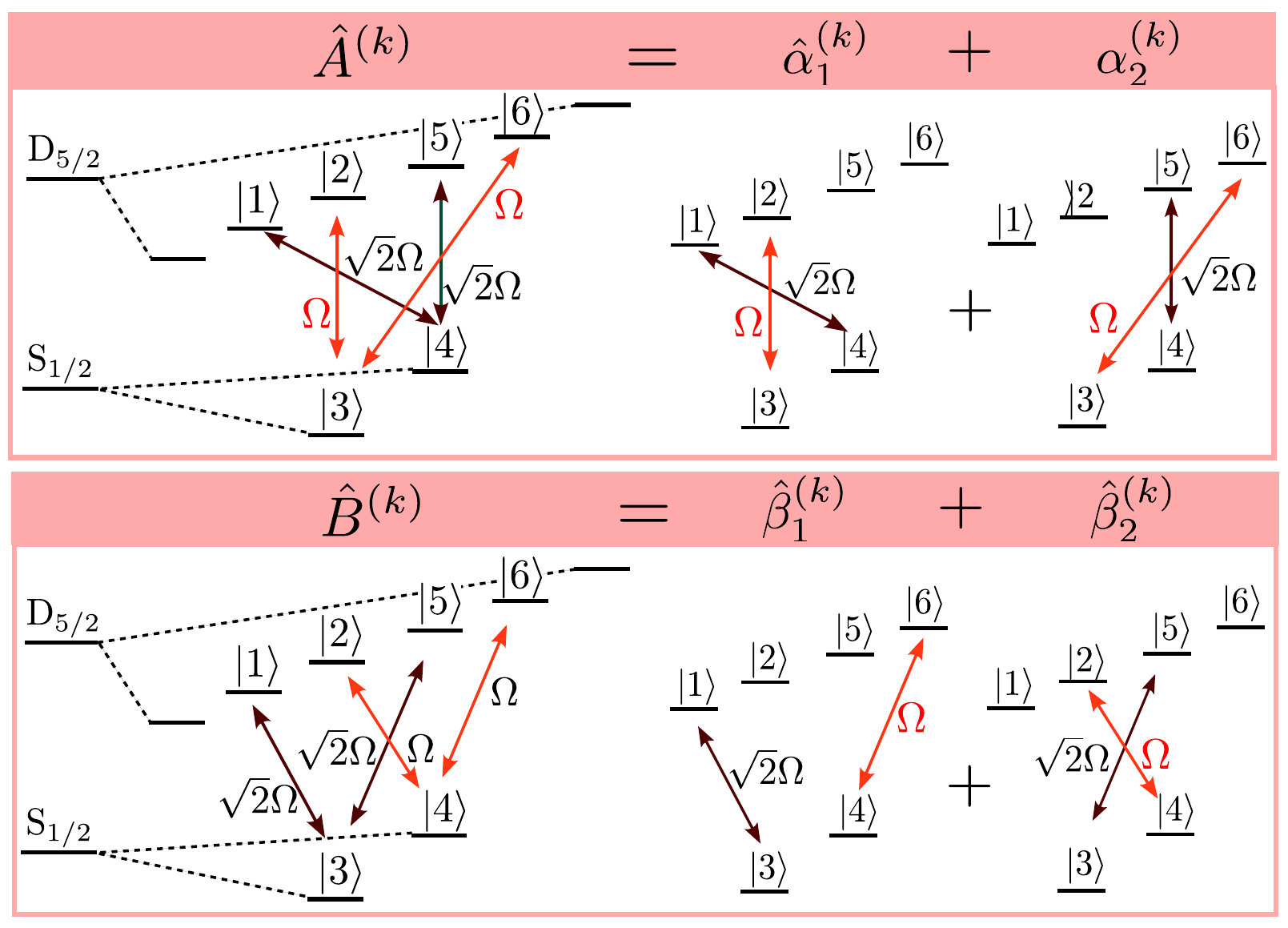}%
    \caption{\textit{Model encoding into ions qudit}.-- The six states needed for the model \eqref{eq:Heff} are encoded within the eight Zeeman sublevels of the $S_{1/2}$ and $D_{5/2}$ states of $^{40}\textrm{Ca}^+$ ions.
     By encoding the states $|3\rangle$ and $|4\rangle$ into the $S_{1/2}$ manifold all matrix elements can be implemented via direct transitions. The boxes on the left show the driving scheme to implement the $\hat A^{(k)}$ and $\hat B^{(k)}$ matrices, $k=1,2$, while the boxes on the right show the implementation of the $\hat \alpha^{(k)}$ and $\hat \beta^{(k)}$ matrices.
     The relative Rabi frequencies needed for each transition are indicated next to the corresponding arrows with two different colors, red for $\Omega$ and dark red for $\sqrt{2}\Omega$ .} \label{fig:level}
\end{figure}

The structure of the truncated gauge-preserving Hamiltonian given in Eq.~\eqref{eq:Heff}, defined on a dressed local basis of dimension six, suggests a natural implementation on a qudit-based quantum processor. In the following, we  focus on an implementation using trapped $^{40}\textrm{Ca}^+$ ions, as presented in Ref.~\cite{ringbauer2022universal}, but the proposed scheme is versatile and applicable also to other qudit quantum processors.
In the trapped-ion experiment of Ref.~\cite{ringbauer2022universal}, each qudit is encoded within the electronic ground state $S_{1/2}$ and the metastable excited state $D_{5/2},$ as illustrated in Fig.~\ref{fig:level}. An external magnetic field splits the $S_{1/2}$ state into two Zeeman sublevels, $m_B=\pm 1/2,$, and the excited state into six Zeeman sublevels ($m_B=\pm 5/2, \pm 3/2, \pm 1/2$).
 This configuration yields a total of eight accessible qudit levels connected, considering selection rules, by ten allowed electric-quadrupole transitions ($\Delta m_B=0,\pm 1, \pm 2$). A convenient encoding of the model given in Eq.~\eqref{eq:Heff} onto the ion's levels involves placing the states $|3\rangle$ and $|4\rangle$ within the $S_{1/2}$ ground state, while the remaining states reside in the $m_B=\pm 3/2$ and $m_B=\pm 1/2$ levels of the $D_{5/2}$ metastable state, as illustrated in Fig.~\ref{fig:level}. This choice is motivated by the observation that all the transitions in the  model are interconnected via the two states: $|3\rangle$ and $|4\rangle$. 
Note that, compared to other encoding strategies~\cite{gonzalez2022hardware,zache2023fermion,popov2023variational}, where each gauge link and each fermion site are separately associated with a qudit, our approach only requires $N$ qudits to describe both gauge and matter fields in a lattice consisting of $N$ sites. Such a reduction in the number of employed atoms is an important advantage for ion-based simulators, where scalability in the number of atoms is a standing challenge.
 
Let us now discuss the implementation of the different  terms composing the qudit Hamiltonian~\eqref{eq:Heff}.
 Generic single qudit operations among these states can be decomposed into at most $d(d-1)/2$ two-level subspace rotations, 
 where $d$ is the qudit dimension. These rotations have the form
 $\hat R(\theta,\phi)=e^{-i(\theta/2)\hat\sigma^{s,s'}_\phi}$, where 
 the operators $\hat\sigma^{s,s'}_\phi=|s\rangle\langle s'|$ connect the qudit states $s$ and $s'$,
 $\theta$ denotes the rotation angle, and $\phi$ sets the rotation axis.
Importantly, in the model defined in Eq.~\eqref{eq:Heff} the mass and gauge Hamiltonians are diagonal, requiring only $5$ elementary rotations, which can be easily implemented with high fidelity.

The hopping part of the Hamiltonian given in Eq.~\eqref{eq:Heff} is more demanding. It necessitates two-qudit operations, which can be broken down into a sequence of entangling gates operating on pairs of qudit levels (we omit the site subscript $n$ for notation simplicity): %By defining the extended Pauli matrices, denoted as $\tilde\sigma^{s,s'}_\nu=|s\rangle\langle s'|$, which  connect the qudit states $s$ and $s'$ via a rotation along the  $\nu$ axis,
%These interaction matrices can be decomposed   in Eq.~\eqref{eq:Heff} as follow:
\begin{equation}
\begin{split}
\hat A^{(1)}&=\hat\sigma^{2,3}_x+\sqrt{2}\hat\sigma^{1,4}_x+\sqrt{2}\hat\sigma^{4,5}_x+\hat\sigma^{3,6}_x\\
\hat A^{(2)}&=-(\hat\sigma^{2,3}_y+\sqrt{2}\hat\sigma^{1,4}_y+\sqrt{2}\hat\sigma^{4,5}_y+\hat\sigma^{3,6}_y)\\
\hat B^{(1)}&=\hat\sigma^{2,4}_y+\sqrt{2}\hat\sigma^{1,3}_y+\sqrt{2}\hat\sigma^{3,5}_y+\hat\sigma^{4,6}_y\\
\hat B^{(2)}&=\hat\sigma^{2,4}_x+\sqrt{2}\hat\sigma^{1,3}_x+\sqrt{2}\hat\sigma^{3,5}_x+\hat\sigma^{4,6}_x\,.
\end{split}
\end{equation}
Note that with the chosen encoding of the model this decomposition involves only direct transitions between the  $S_{1/2}$ and $D_{5/2}$
states.
From this decomposition, it becomes evident that a single next-neighbor ($n,n+1$) block of the hopping Hamiltonian requires a total of $32$  entangling gates %two-qubit gates, %($64$ for a unit circuit cell made of three lattice sites), 
of the form $\hat R_{XY}(\varphi)=e^{-i\varphi\hat\sigma^{s_1,s_2}_x \otimes \hat\sigma^{s_3,s_4}_y}$. %or $R_{YX}$. 
These rotations can be implemented using M{\o}lmer S{\o}rensen (MS) gates~\cite{molmer1999multiparticle}. 
Although this decomposition requires resources available on the current trapped ion qudit quantum processor~\cite{ringbauer2022universal},
 the considerable gate count imposes limitations on the performance of a quantum simulation of the model. In the following, we will explore how this scheme can be enhanced by using native qudit gates based on the simultaneous driving of multiple transitions.

 %large gate count, which can be drastically improved by making use of native qudit gates based on a simultaneous driving of multiple transitions.

%The previous decomposition has shown that, to implement  a single evolution time step induced by the interaction Hamiltonian involving two ions, 32 MS gates are required. 
%Such requirement is at the moment  way beyond the state of the art. Indeed in Ref.\cite{ringbauer2022universal} is clearly shown that already after 10 MS gates the Fidelity of the operation drops below $80\%$.  To see how to improve this scheme  we present in the following section an alternative decomposition involving native qudit gates.

%\subsection{Two-qubit gates decomposition}

\subsection{Native two-qudit gates}
The two hopping blocks, denoted as $\hat A^{(k)}_n\hat B^{(k)}_{n+1}$ with $k=1,2$, can be directly implemented through a generalized M{\o}lmer-S{\o}rensen scheme~\cite{low2020practical} that involves the simultaneous driving of all four direct transitions in each ion, as depicted in Fig.~\ref{fig:level}.
Similar to standard MS gates, assume that each transition of interest, $\omega_{s,s'}$, is driven by a pair of lasers with frequencies $\omega^1_L$ and $\omega^2_L$, featuring opposite detunings, i.e., $\omega^1_L=\omega_{s,s'}+\delta$ and $\omega^2_L=\omega_{s,s'}-\delta$. To achieve the correct matrix elements reported in Tab.~\ref{tab:effmatrices1}, two out of the four transitions are driven with a Rabi frequency of $\Omega$, while the other two are driven with $\sqrt{2}\Omega$, see Fig.~\ref{fig:level}. Additionally, the phases $\phi_n$ associated with each ion's driving are chosen to align with the correct phase pattern.
We assume, to work in the Lamb-Dicke regime, $\eta\ll 1$, with $\eta$ being the Lamb-Dicke parameter, 
and to separate the blue and red sidebands, coming from each laser pair, via a rotating wave approximation valid in the regime where the lasers are tuned close to the motional sideband with frequency $\nu$, i.e., $|\delta-\nu|\ll \delta$.
A hopping block for two target ions $n$ and $n+1$  can then be implemented via the generalized MS Hamiltonian
\begin{equation}\label{eq:GMS}
\hat {\bar H}^{(k)}_{\rm \hat A_n \hat B_{n+1}}=\hbar\frac{\eta\Omega}{2}\left[\hat A^{(k)}_{n}+\hat B^{(k)}_{n+1}\right][\hat a^{\dagger}e^{i(\nu-\delta)t}+\hat ae^{-i(\nu-\delta)t}],
\end{equation}
where $\hat a$ ($\hat a^\dagger$) is the phonon annihilation (creation) operator. In the regime of weak driving, $\eta\Omega\ll|\nu-\delta|$, the phononic bath dispersively mediates interactions among the two ions according to the effective Hamiltonian 
\begin{equation}\label{eq:HMSeff2}
\hat {\bar H}^{(k)}_{\rm \hat A_n \hat B_{n+1}}\simeq\hbar\frac{(\eta\Omega)^2}{4(\nu-\delta)}\left[\hat A^{(k)}_{n}+\hat B^{(k)}_{n+1}\right]^2,
\end{equation}
where, in order to minimize the population of the motional mode, we assumed to let the system evolve for a time $d \bar{t}=2\pi \ell /|\nu-\delta|$, with $\ell$ being a positive integer~\cite{molmer1999multiparticle,low2020practical}.
To perform the dynamics in natural units, as in Eq.~\eqref{eq:Heff}, we re-scale this time by the rate associated with the MS gate, i.e. $d \bar{t}\rightarrow dt = \pi(\eta\Omega/|\nu-\delta|)^2$. %{\color{red} I anticipated here the rescaling discussion }
The Hamiltonian~\eqref{eq:HMSeff2}, once re-scaled, exactly reproduces the desired hopping block up to unwanted single qudit rotations coming from the terms $(\hat A^{(k)}_{n})^2$ and $(\hat  B^{(k)}_{n+1})^2$. Upon aggregating contributions from all transitions, these rotations simplify to straightforward diagonal matrices. 
These terms can then be combined with the mass and gauge Hamiltonians of Eq.~\eqref{eq:Heff}, resulting in the following single qudit Hamiltonian:
\begin{equation}\label{eq:Hind}
\hat H_{n}= m(-1)^{n}\hat M_n+g^2\hat C_n-\hat H_n^{A^2}-\hat H_n^{B^2}
\end{equation}
where the diagonal terms in natural units read
\begin{equation}
\begin{split}\label{eq:corr}
&\hat H_n^{A^2}={\rm diag}(2,1,2,4,2,1)\\
&\hat H_n^{B^2}={\rm diag}(2,1,4,2,2,1)\,.
\end{split}
\end{equation}
Note that, for $n=1$, besides the gauge and mass terms, only $\hat H_n^{A^2}$ contributes to Hamiltonian~\eqref{eq:Hind}, while for $n=N$ only the $\hat H_n^{B^2}$ term does. 
The Hamiltonian terms given in Eq.~\eqref{eq:HMSeff2} and Eq.~\eqref{eq:Hind} constitute the fundamental building block of the digital quantum simulation to be discussed in Section~\ref{Sec.Dig_sim}. Importantly, with this procedure involving generalized MS gates based on the simultaneous driving of four transitions, only 2 entangling operations (one per each hopping block) are necessary to implement the hopping between two neighboring sites.

\begin{figure*}[!t]
    \includegraphics[width=1.0\textwidth]{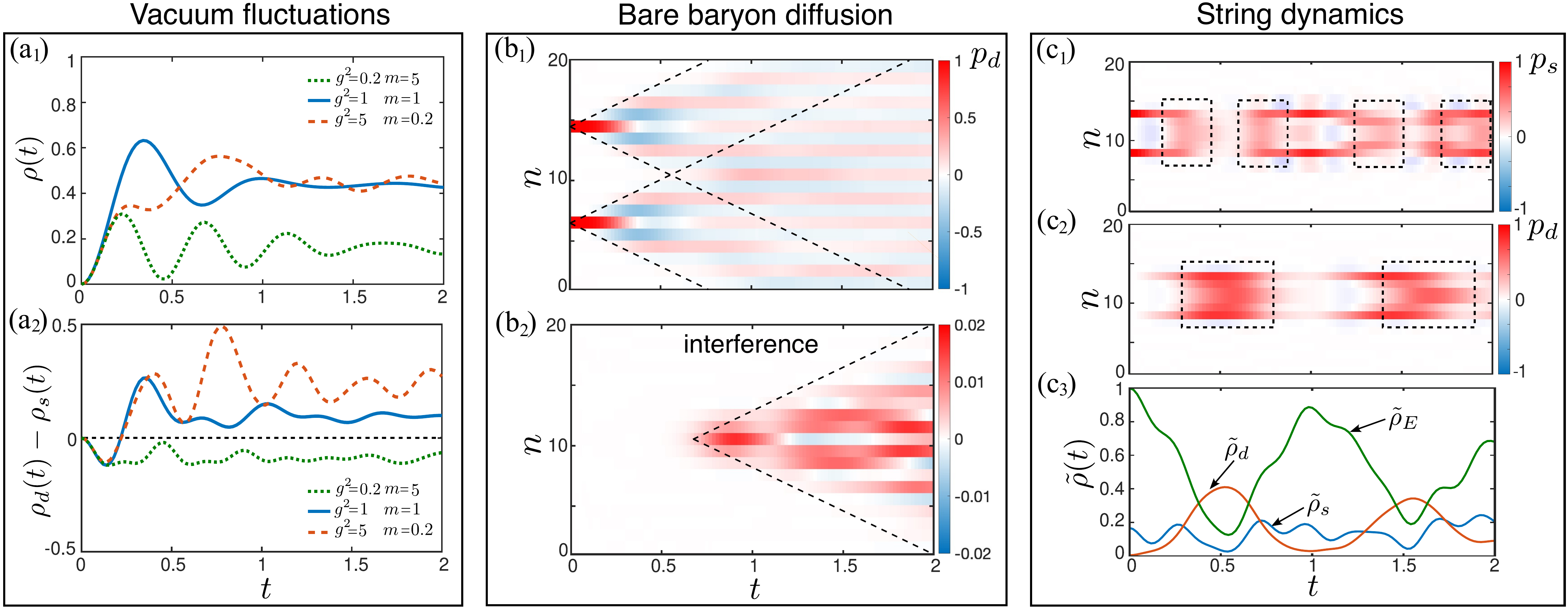}%
    \caption{\textit{Model phenomenolgy}.--(a) Pair production. Particle density ($\rm a_1$) and the difference between double and single particle occupancy densities ($\rm a_2$) as a function of time for different mass-coupling values as indicated in the figure. (b) Bare baryon diffusion. ($\rm b_1$) Double occupancy probability distribution as a function of time capturing the evolution of two bare baryons initially excited at positions $n=6$ and $n=14$ with $g^2=m=0.5$. To better resolve the dynamics we subtracted the  probability distribution associated to the evolution of the vacuum state. ($\rm b_2$) The interference pattern of the process described in ($\rm b_1$) obtained by subtracting the individual diffusion of each baryon.  (c) String dynamics. Single ($\rm c_1$) and double ($\rm c_2$) occupancy probability distributions as a function of time for a quark-antiquark string initially excited within the sites $n=8-13$ with $g^2=m=5$. Here we again subtracted the vacuum state evolution.
     ($\rm c_3$) Electric field and single and double occupancy string densities as a function of time for the same scenario as the two panels above.
   All the simulations have been carried out for a lattice of $N=20$ sites solving the dynamics of the model ~\eqref{eq:Heff} with matrix product states setting the truncation tolerance to $\rm tol = 10^{-7}$ and the maximum bond dimension to $\rm D_{\rm max} = 200$.}
    \label{fig:pheno}
\end{figure*}

\subsection{Intermediate scheme}
The native qudit-gate scheme outlined in the previous section, offers the advantage of achieving an exceptionally short circuit depth. Nevertheless, the implementation of this method presents technical challenges, primarily stemming from the requirement for fine-tuned calibration of all driven transitions, see Sec.~\ref{Sec.experiment}. Fortunately, this demand can be released by an intermediate scheme relying on the simultaneous driving of only two distinct  disjoint transitions. The core idea involves decomposing the interaction matrices as $\hat A^{(k)}=\hat \alpha^{(k)}_1+\hat \alpha^{(k)}_2$ and $\hat B^{(k)}=\hat \beta^{(k)}_1+\hat \beta^{(k)}_2$, where $k=1,2$ and
\begin{equation}
\begin{split}
\hat \alpha^{(1)}_1&=\hat\sigma^{2,3}_x+\sqrt{2}\hat\sigma^{1,4}_x
\qquad\qquad\quad
\hat \alpha^{(2)}_1=-(\hat\sigma^{2,3}_y+\sqrt{2}\hat\sigma^{1,4}_y)\\
\hat \alpha^{(1)}_2&=\hat\sigma^{3,6}_x+\sqrt{2}\hat\sigma^{4,5}_x
\qquad\qquad\quad
\hat \alpha^{(2)}_2=-(\hat\sigma^{3,6}_y+\sqrt{2}\hat\sigma^{4,5}_y)\\
\hat \beta^{(1)}_1&=\hat\sigma^{4,6}_y+\sqrt{2}\hat\sigma^{1,3}_y
\qquad\qquad\quad
\hat \beta^{(2)}_1=\hat\sigma^{4,6}_x+\sqrt{2}\hat\sigma^{1,3}_x\\
\hat \beta^{(1)}_2&=\hat\sigma^{2,4}_y+\sqrt{2}\hat\sigma^{3,5}_y
\qquad\qquad\quad
\hat \beta^{(2)}_2=\hat\sigma^{2,4}_x+\sqrt{2}\hat\sigma^{3,5}_x
\end{split}\,,
\end{equation}
  is one of the possible decompositions involving only disjoint transitions. The following discussion still holds for other possible decomposition choices.
With this scheme, in contrast to the earlier approach involving only two entangling gates, a total of 8 operations are required to reproduce the  hopping between two neighboring sites. The different contributions can be once again implemented by simultaneously driving the two target transitions with a pair of bi-chromatic pulses characterized by Rabi frequencies ($\Omega, \sqrt{2}\Omega$), as illustrated in Fig.~\ref{fig:level}. Assuming the validity of the same assumptions employed in the preceding section, we derive the effective Hamiltonian
\begin{equation}\label{Halphabeta}
\hat{\bar H}^{(k)}_{\rm \hat\alpha_{q,n}\hat\beta_{q',n+1}}\simeq\hbar\frac{(\eta\Omega)^2}{4(\nu-\delta)}\left[\hat \alpha^{(k)}_{q,n}+\hat \beta^{(k)}_{q',n+1}\right]^2,
\end{equation}
where $q,q'=1,2$ and we again set the time step to $d\bar 
t=2\pi \ell /|\nu-\delta|$. 
As before, the phonon-mediated interaction induces
unwanted single qudit transitions that reduce, after re-scaling in the natural units,  to twice the same diagonal matrices of in Eq.~\eqref{eq:corr}, $\hat H_{n}= m(-1)^{n}\hat M_n+g^2\hat C_n-2\hat H_n^{A^2}-2\hat H_n^{B^2}$.% $\sum_{k,q}(\hat \alpha^{(k)}_{q,n})^2\equiv\sum_{k}(\hat A_n^{(k)})^2$, being
%$\sum_k(\hat \alpha^{(k)}_{1,n}\hat \alpha^{(k)}_{2,n}+\hat \alpha^{(k)}_{2,n}\hat \alpha^{(k)}_{1,n})=0$, with  the same holding for the $\beta^{(k)}_{q,n}$ matrices.
%These contributions  can be   compensated by performing a single qudit rotation on each ions at the end of each time step. The  circuit decomposition for a time step evolution of the sites $n$ and $n+1$ according to this scheme is shown in Fig.~\ref{fig:circuit}(b). 
%We finally observe that the convenience of this scheme relies on the fact that, for all the decomposed terms, the two simultaneously driven transitions never involve the same ground state. This property should strongly reduce the calibration issues compared to the first proposed scheme.

\section{Model phenomenology}\label{Sec.phen}
In this section we consider a few paradigmatic examples of non-trivial dynamics occurring in the model by evolving the initial state $|\Psi(t=0)\rangle$ under the time-dependent Schr\"{o}dinger equation ruled by the Hamiltonian given in Eq.~\eqref{eq:Heff}. In particular, we identify phenomena and observables where the non-abelian nature of the theory manifests itself with distinctive features and that could be implemented in current state of the art experiments.

\subsection{Vacuum fluctuations}
We first consider  the phenomena of particle density fluctuations starting from an initial false vacuum~\cite{banerjee2012atomic,kuhn2014quantum,magnifico2020real}. Let us assume the system to be initially in the Dirac sea, i.e. the bare vacuum represented by the ground state of the free part of Hamiltonian~\eqref{eq:Heff} for large positive mass and coupling, which consists in alternated empty and doubly occupied quark and anti-quark sites, $|\Psi(t=0)\rangle=| 5\rangle| 1\rangle...| 5\rangle| 1\rangle$. When the hopping turns on, this state does not represent the real ground state anymore and it undergoes a non-trivial dynamical evolution: depending on the Hamiltonian parameters, %quantum fluctuations can stimulate the 
spontaneous production of quark anti-quark pairs out of the vacuum takes place. 
This effect can be quantified by the lattice particle density counting for the total number of created quark and anti-quark excitations, %which we compute from the  population on each state $s$ at the site $n$, 
\begin{equation}\label{eq:fermion_density}
\rho(t)=\rho_{s}(t)+\rho_{d}(t)\,,
\end{equation}
where $\rho_{e}(t)=1/(2N)\sum_{n}p_{e}(n,t)$ represent respectively the single $(e=s)$ and double $(e=d)$ particle occupancy densities computed from the probability distributions:
\begin{equation}
p_{s}(n,t)=|\langle \Psi(t)|3    \rangle_n|^2+|\langle \Psi(t)|4    \rangle_n|^2\,,
\end{equation}
and  
\begin{equation}
p_d(n,t)=2\begin{cases}
             |\langle \Psi(t)|5  \rangle_n|^2+|\langle \Psi(t)|6    \rangle_n|^2 \;\;\;\;\text{if $n\in \rm even$} \\
                        |\langle \Psi(t)|1  \rangle_n|^2+|\langle \Psi(t)|2    \rangle_n|^2 \;\;\;\;\text{if $n\in \rm odd$}\,.
                    \end{cases}
\end{equation}
In Fig.~\ref{fig:pheno}($\rm a_1$) we simulated the evolution of the particle density for different mass-coupling ratios starting from the Dirac %vacuum 
sea. % where there are neither quark nor anti-quark excitations. When the interaction is turned on the system moves away from the initial ground state and produces particles in pairs. 
For small masses, $m\ll 1$, particle production is energetically favorable, and after a transient the system reaches a steady state with approximately one particle per lattice site. For large masses, $m\gg 1$, instead the particle production is more expensive and recombination effects let the system oscillate between the Dirac and the true vacuum. %real vacuum. 
To probe the non-abelian nature of the model we compute separately the contributions to the particle density coming from the single and double particle occupancy densities. In turn, these quantities encode the population of bare excitations such as color-singlet pairs formed by a quark and an anti-quark occupying two neighboring sites with an excited shared gauge link (bare meson), and quark-quark (bare baryon) or anti-quark--anti-quark (bare anti-baryon) pairs occupying the same site. 
%are related to the meson (quark--anti-quark pair) and baryon (quark--quark or anti-quark--anti-quark pair) populations.
 Note that the  baryon population %the  with the second being 
is absent in abelian models where just one particle can be hosted in each lattice site.  
To quantify the different weights of the two contributions to the total particle density we plot in Fig.~\ref{fig:pheno}($\rm a_2$) the difference between double and single particle occupation densities
as a function of time for the same mass-coupling ratios as in Fig.~\ref{fig:pheno}($\rm a_1$).
The figure shows how single particle occupation is dominant in the weak coupling regime, $g^2\ll 1,m$, %\danger{[wait, why $g^2\ll m$ ? Shouldn't it be $g^2\ll 1$ ?]}
 while in the strong coupling regime $g^2\gg 1,m$ %\danger{[again $g^2\gg m$ I think?Si esattamente  per $g^2\gg 1,m$ ]} 
the production of  bare baryons is more energetically favorable. %are produced via a second order process at an effective rate of $J_{\rm eff}=4/g^2$. 
%\danger{[Veramente mi risulta che sia
%$4(g^2 + 2m)^{-1}(g^2 + 4m)^{-1}$ a parte un fattore moltiplicativo Reply: lo ricontrollo.]}
Such  baryon production has no analogs in abelian theories and is a clear signature of the underlying $SU(2)$ nature of the model.

\begin{figure*}[!t]
    \includegraphics[width=0.95\textwidth]{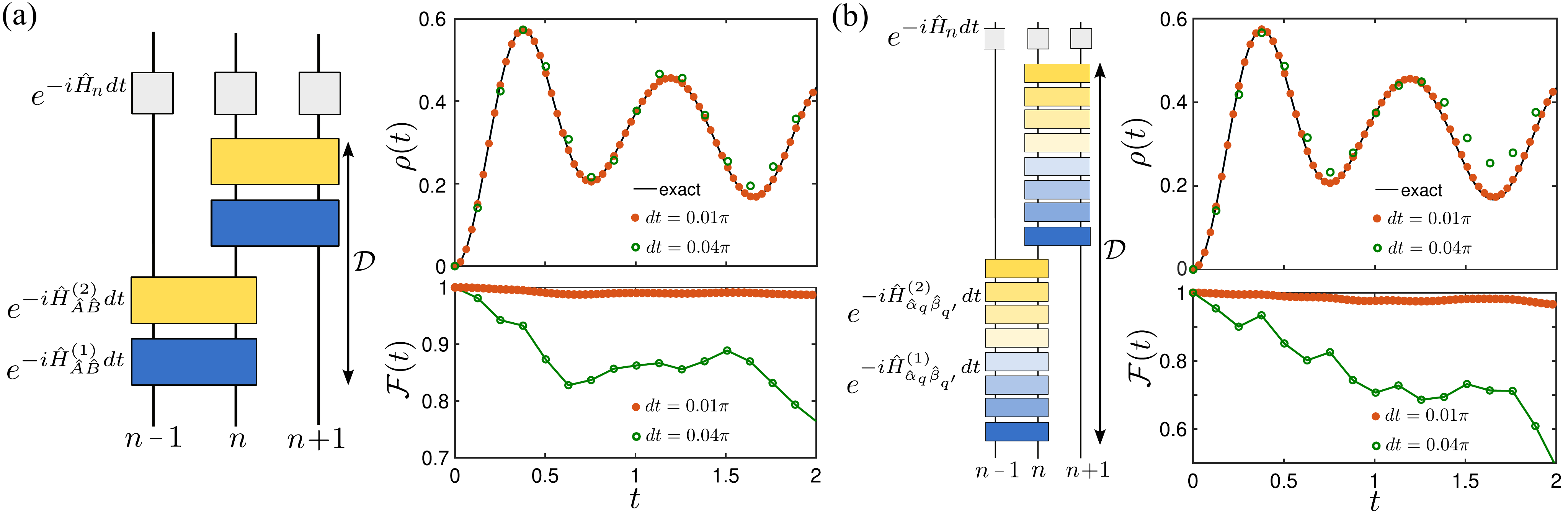}%
    \caption{\textit{Digital quantum simulation}.--(a)-(b) Left panels: Circuit decomposition for the two strategies outlined in the main text, (a) full and (b) intermediate scheme. In both $\mathcal{D}$ indicates the circuit depth.
    Right panels: Dynamics of the particle density evolving from the Dirac vacuum for $N=3$ lattice sites obtained through the model~\eqref{eq:Heff} (continuous lines) and via a first-order Suzuki-Trotter evolution of the illustrated circuits. In the latter, we explicitly included the phononic mode as in Eq.~\eqref{eq:GMS} truncating at the eighth level. The corresponding fidelity between the exact and the digitally simulated state is also shown. The time steps used for the Suzuki-Trotter evolution are indicated in the figures. %and correspond respectively to the Rabi frequency-detuning ratios: $\eta\Omega/|\nu-\delta|=0.1$ and $\eta\Omega/|\nu-\delta|=0.2$.
    In all the simulations, we set  $m=g^2$. %and we re-scaled the time to natural unit: $t:=t(\eta\Omega)^2/(2|\nu-\delta|)$ and $dt:=dt\pi(\eta\Omega/|\nu-\delta|)^2$ .
   }\label{fig:circuit}
\end{figure*}

\subsection{Bare baryon diffusion}
The second dynamical phenomenon under consideration involves the diffusion of two pairs of quarks (bare baryons) initially localized on two even sites of the lattice, with the rest of the system being in  the Dirac vacuum.
%initially localized at an empty site of the Dirac vacuum. 
In the strong coupling limit, where $g\gg m,1$, these pairs play the role of composite particles in the lattice, hopping from one matter site to another, weakly exciting all other allowable configurations.
This regime, characterized by a fourth-order process with a rate of $J_{\rm eff}=16/g^4m$, 
%\danger{[Non mi torna, dovrebbe esserci un $g^{-8}$ se siamo al quarto ordine]}{\color{red}(GM: Dove viene presentata questa descrizione?)}, 
occurs on a timescale significantly longer than the natural unit (see App.\ref{App.Bar_diff}). Consequently, it is not practically feasible for digital quantum simulation due to the requirement for numerous time steps (see Sec.\ref{Sec.Dig_sim}).

To investigate baryon diffusion on a shorter timescale, we examine the non-perturbative regime where $g\sim m\lesssim 1$. Specifically, we focus on the scenario where two bare baryons are initially excited out of the Dirac vacuum, $\hat S^{51}_{n_1}\hat S^{51}_{n_2}|{\rm GS}\rangle$, where $\hat S^{ss'}_{n}=|s\rangle\langle s'|_{n}$ with $n_1$ and $n_2$ being even numbers.
In Fig.~\ref{fig:pheno}($\rm b_1$), the evolution of the double occupancy probability distribution, $p_d$, is plotted, with the vacuum fluctuation subtracted to reveal the diffusion process. The two bare baryons disperse within a wavefront cone with an aperture of approximately $\pm 8t$, governed by the quark hopping rate.
The interference of the two bare baryons is depicted in Fig.~\ref{fig:pheno}($\rm b_2$) and is obtained by subtracting the individual bare baryon propagations. %It's noteworthy that this observed dynamics lacks an analog in abelian models where double-site occupancy is forbidden. 

\subsection{String dynamics}
%The last phenomenon we consider is string-breaking,
%a characteristic of LGTs in confined phases. In this sce-
%nario, an initial string excitation undergoes evolution by
%breaking into pairs of particles and antiparticles.
Finally, we  consider  the dynamics of an initially excited string. In this context, string-breaking is a typical phenomenon occurring in confined LGTs where
 an initial string excitation undergoes evolution by breaking into pairs of particles and antiparticles. This phenomenon has been extensively studied numerically in (1+1)D U(1) models~\cite{pichler2016real,magnifico2020real,zhang2023observation,mildenberger2022probing} and in a (1+1)D SU(2) theory using a different rishon representation than the one employed here~\cite{kuhn2015non}.
To initialize the state configuration, we apply the following string operator of length $l$ to the Dirac sea vacuum: $\mathcal{\hat S}=\hat\psi^{\dagger}_{n_s} \hat U_{n_s,n_s+1},...\hat U_{n_s+l-1,n_s+l}\hat\psi_{n_s+l}$ where we assume $n_s$ to be even, corresponding to the creation of a  quark, and $n_s+l$ to be odd, corresponding to the creation of an anti-quark on that site. Expressed in the local dressed basis of Eq.~\eqref{eq:dressed basis} this string reads:
\begin{equation}
\mathcal{\hat S}=\hat S^{41}_{n_s}\hat S^{65}_{n_s+1}\hat S^{21}_{n_s+2}\;...\;\hat S^{65}_{n_s+l-2}\hat S^{21}_{n_s+l-1}\hat S^{35}_{n_s+l}\,.
\end{equation}
%the following string operator of length $l$, $\mathcal{\hat S}=\hat\psi^{\dagger}_{n_s} \hat U_{n_s,n_s+1},...\hat U_{n_s+l-1,n_s+l}\hat\psi_{n_s+l}$
%To initialize the state configuration, we apply the following string operator of length $l$ to the Dirac sea vacuum:
%\begin{equation}
%\mathcal{\hat S}=\hat S^{n_s}_{41}\hat S^{n_s+1}_{65}\hat S^{n_s+2}_{21}\;...\;\hat S^{n_s+l-%2}_{65}\hat S^{n_s+l-1}_{21}\hat S^{n_s+l}_{35}\,,
%\end{equation}
%where we assume $n_s$ to be even, corresponding to the creation of a  quark, and $n_s+l$ to be odd, corresponding to the creation of an anti-quark on that site. 
%It's worth noting that this string corresponds to $\mathcal{\hat S}=\hat\psi^{\dagger}_{n_s} \hat U_{n_s,n_s+1},...\hat U_{n_s+l-1,n_s+l}\hat\psi_{n_s+l}$ when expressed in terms of the original model operators of Eq.~\eqref{eq:H}.

Here we focus again on distinctive non-abelian features by distinguishing between bare meson and baryon resonant production from the string.
To observe resonant production of bare baryons and mesons, we consider the case where the initial string of energy $E_{\rm str}=2m+2lg^2$ resonates with the energy of $l+1$ baryons, each having energy $E_{\rm bar}=2m$, and $(l+1)/2$ mesons, each having energy $E_{\rm mes}=2m+2g^2$. This resonance condition is fulfilled for both processes when $g^2=m$. We then let the system evolve in time with fixed mass and coupling set to large values to ensure that most of the energy remains confined within the string, preventing quick dispersion toward the system edges. 
In Fig.~\ref{fig:pheno}($\rm c$) we indeed observe resonant excitations in time of the single and double occupancy probability distributions within the string, signaling the creation of the string of bare baryons and mesons. These resonant oscillations are better resolved in Fig.~\ref{fig:pheno}($\rm c_3$) where we plot the single $(e=s)$ and double $(e=d)$ occupancy densities averaged within the string, $\tilde \rho_{e}(t)=1/(2(l+1))\sum^{n_s+l}_{n=n_s}p_{e}(n,t)$, along with the electric field string density defined as
\begin{equation}
\begin{split}
 & \tilde \rho_{E}=\frac{1}{2l}\sum^{n_s+l-1}_{n=n_s} |\langle \Psi(t)|2  \rangle_n|^2+ |\langle \Psi(t)|4  \rangle_n|^2+ |\langle \Psi(t)|6  \rangle_n|^2 \\
 &+|\langle \Psi(t)|2  \rangle_{n+1}|^2+|\langle \Psi(t)|3  \rangle_{n+1}|^2+|\langle \Psi(t)|6  \rangle_{n+1}|^2\,,
\end{split}
\end{equation}
which accounts for the number of rishons on each link.

\section{Digital quantum simulation}\label{Sec.Dig_sim}

In this section, we explore how the dynamics previously described can be simulated digitally using a trapped-ion qudit quantum processor. 
To digitally simulate the dynamics, we use the first-order Suzuki-Trotter decomposition:
\begin{equation}\label{Eqtrotter}
 \mathcal{\hat U}(t_{\rm f})=e^{-i\hat H t_{\rm f} }\simeq\left(\prod_j e^{-i\hat h_j dt}\right)^{n_{\rm ST}}\,,
\end{equation}
where, $dt=t_{\rm f}/n_{\rm ST}$ denotes the time step ruling the precision of the  Suzuki-Trotter evolution, $n_{\rm ST}$ is the number of Suzuki-Trotter steps, and $t_{\rm f}$ is the final time of the evolution. The terms $h_j$ represent the Hamiltonian building blocks composing Hamiltonian~\eqref{eq:Heff}, which we presented in Sec.~\ref{Sec:encoding}.
%we indicate the various terms that compose the Hamiltonian \eqref{eq:Heff}, and their implementation has been detailed in Sec.~\ref{Sec:encoding}. 
%The evolution of the model is then obtained by executing in parallel for each time step the unit circuit cell,
%involving three lattice sites, depicted in Fig.~\ref{fig:circuit} for both schemes presented in Sec.~\ref{Sec:encoding}. 
The evolution of the model is then obtained by executing, for each time step, the unit circuit cell (in parallel across the lattice), depicted in Fig.~\ref{fig:circuit} for both schemes presented in Sec.~\ref{Sec:encoding}.
The unit circuit cell involves three lattice sites and is composed of the generalized MS gates, implementing the hopping part of the model, followed by a single-qudit rotation on each ion encompassing the gauge and mass terms of the Hamiltonian~\eqref{eq:Hind}, and the correcting rotations. Considering that single-qudit rotations can be performed with high fidelity, we define the circuit depth of the model, $\mathcal{D}$, as the number of generalized MS gates per unit circuit cell. 
Then the total number of two-qudit gates, $N_{\rm gates}$, required to perform the entire simulation on a chain of $N$ sites scales linearly with the system size and the Trotter steps:
 \begin{equation}\label{eq:ngate}
 N_{\rm gates}=\frac{(N-1)}{2}n_{\rm ST}\mathcal{D},
 \end{equation}
with $\mathcal{D}=4$ and $\mathcal{D}=16$ respectively for the first and second scheme presented in Sec.~\ref{Sec:encoding}. Note that the total gate count of our proposal for both schemes is drastically reduced with respect to a computation based on a decomposition into two-level entangling gates (see Sec.~\ref{Sec:encoding}), which has $\mathcal{D}=64$. 
%Finally, a single qudit rotation for each site is performed to compensate for unwanted rotations (see Sec.~\ref{Sec:encoding}). 

\begin{figure*}[!t]
    \includegraphics[width=1\textwidth]{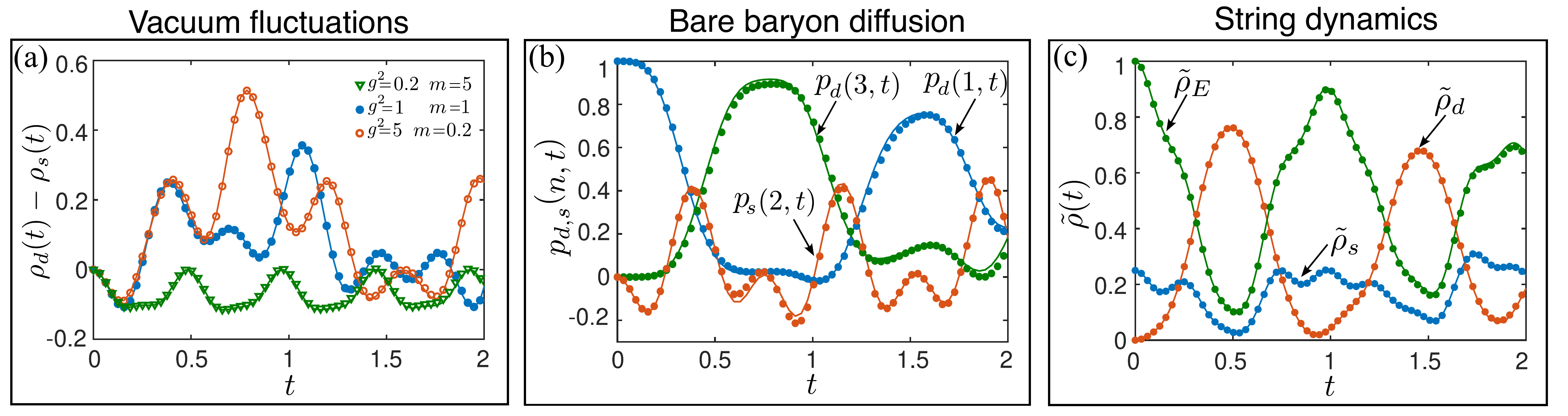}%
    \caption{\textit{Reproducing model phenomenology}.--(a) Difference between double and single particle occupancy densities as a function of time for a lattice with $N=3$ sites, considering various mass-coupling values as depicted in the figure.
  (b) Bare baryon diffusion is analyzed for a lattice with $N=3$ sites and $g^2=m=0.5$. The probabilities of double occupancy for the first and third atoms, denoted as $p_d(1,t)$ and $p_d(3,t)$, along with the single occupancy probability for the second atom, denoted as $p_s(2,t)$, are studied over time. The initial condition involves an bare anti-baryon excited in the first site.
(c) Electric field, as well as single and double occupancy string densities, are plotted over time for a lattice comprising $N=4$ sites with $g^2=m=5$. The string is initially excited within the sites $n=1-4$. 
In all plots, continuous lines represent the exact evolution of the model~\eqref{eq:Heff}, while discrete points correspond to the first-order Suzuki-Trotter evolution, with time step $dt=0.01\pi$, performed using~\eqref{eq:GMS}, which includes 8 levels of the phononic mode.
   }\label{fig:sim_fewsites}
\end{figure*}

For a realistic numerical simulation of the model, we consider small lattice sizes, specifically $N=3$ and $N=4$. These sizes are chosen as they retain the essential characteristics of the SU(2) dynamics under consideration while being accessible with a short circuit implementation. The generalized MS gates are simulated explicitly including the vibrational degree of freedom using Hamiltonian~\eqref{eq:GMS} and its analogue for the intermediate scheme. %To perform the dynamics in natural units, as in Eq.\eqref{eq:Heff}, we re-scale the physical lab time $\bar t$ by the quark hopping rate, which is expressed in terms of the Rabi frequency and detuning of the pulses, $t=\bar t(\eta\Omega)^2/(2|\nu-\delta|)$. 
%The Suzuki-Trotter time steps are instead conditioned by the necessity of maintaining a negligible phononic population, a requirement met when  $d\bar t=2\pi /|\nu-\delta|$~\cite{molmer1999multiparticle}, which after re-scaling in natural units reads $dt=\pi(\eta\Omega/|\nu-\delta|)^2$. The precision of the digital simulation is then governed by the ratio $(\eta\Omega)/|\nu-\delta|$.
 %In the following, we will describe how this time evolution protocol can be applied to simulate a pair production process where the SU(2) nature of the model becomes evident even with a small number of lattice sites.

\subsection{Reproducing the system dynamics}
To assess the performance of the proposed quantum digital simulation, we compare the results of the exact expected dynamics with those derived from the digital scheme. We apply this comparison to the same illustrative examples of dynamics discussed in Section~\ref{Sec.phen}.
As a first example, we consider in Fig.~\ref{fig:circuit} the particle density evolution of the  Dirac vacuum on $N=3$ lattice sites for the two schemes outlined in Sec.~\ref{Sec:encoding}.
To perform  a comparison  with the exact evolution, we also compute the state fidelity, 
 $\mathcal{F}(t)=\langle\Psi(t)|\hat \varrho_{\rm q}( t)|\Psi(t)\rangle$, where 
 $\hat \varrho_{\rm q}(t)$ represents the reduced density operator of the qudit system after tracing over the phononic degrees of freedom, and $|\Psi( t)\rangle$
is the state evolved under the Schr\"odinger equation.
The two schemes exhibit comparable performances with the precision of the simulation ruled by the Suzuki-Trotter step $dt$. Importantly, even for large time steps, $dt=0.04\pi$, the initial density peak, arising from particle production and characterizing the response time of the system, can be accurately approximated with just $n_{\rm ST}=3$ Suzuki-Trotter steps. Higher simulation precision can be reached by performing a second-order Suzuki-Trotter decomposition, as presented in App.~\ref{App.ST2} and discussed in Sec.~\ref{Sec.experiment}.

Despite the limited lattice size, the digital simulation effectively captures distinct production rates for paired (doubly occupied sites) and unpaired (singly occupied sites) particles.  This is exemplified in Fig.~\ref{fig:sim_fewsites}(a) where we observed a similar phenomenology as the one depicted in  Fig.~\ref{fig:pheno}(a) for the disparity between double and single particle occupation densities.
In addition to particle production from the vacuum, the digital simulation, when confined to a few lattice sites, successfully replicates the other two phenomena discussed in Section~\ref{Sec.phen}: bare baryon diffusion and string dynamics. The former is illustrated in Fig.~\ref{fig:sim_fewsites}(b) where we considered the hopping of an bare anti-baryon excited on the first site towards the third site.   This process is quantified by observing the double particle occupancy probability on the first lattice site, involving an intermediate single particle population on the second site. To enhance the resolution of this effect, we subtracted the Dirac vacuum evolution from the simulation, consistent with the approach taken in Section~\ref{Sec.phen}. The final phenomenon under consideration is string breaking. For this scenario, we extend our analysis to a slightly larger system size, specifically $N=4$. Proceeding as in Fig.~\ref{fig:pheno}($\rm c_3$), we calculate the electric field, as well as the densities of single and double occupancy strings, for a string initially spanning the entire lattice.
As in Fig.~\ref{fig:pheno}($\rm c_3$) we set the string energy to be in resonance with bare meson and baryon production. %This tuning leads to pronounced oscillations in these densities, serving as indicators of the resonant excitations of these particles.
This tuning leads to pronounced oscillations in the quark and  electric field densities, serving as indicators of the resonant excitations of bare mesons and baryons.

\subsection{Link parity preservation and post-selection}\label{Sec.post}
As discussed in Sec.~\ref{Sec.Model}, the employed rishon representation introduces an extra $\mathbb{Z}_2$ symmetry, which we set to an even number of rishons per link. While this symmetry is maintained throughout the Hamiltonian evolution, it may be compromised in actual simulations due to experimental errors leading to leakage from this subspace. These errors can be compensated in post-selection by rejecting states that violate the symmetry, leading to a potentially significant reduction in simulation errors. % by conducting data analysis after the readout, allowing for the selection of states to retain and discard. The benefit lies in the improvement of overall data statistics by rejecting a subset of measurements.

To select which data to retain we make use of the parity matrices $\hat D^{(L)}$ and $\hat D^{(R)}$ defined in Table~\ref{tab:effmatrices2}, which ensures the link fermion parity selection rule.
%\begin{equation}\label{parity}
%\begin{split}
%&\hat P^{(1)}_j =\small
%\begin{pmatrix}
%   1      &  &  & & & \\
%         & -1  &  & & & \\
%    &   & 1 & & & \\
% &   &  & -1 &  & \\
%     &   &  &  & 1 & \\
%     &   &  & & & -1\\
%\end{pmatrix}\\
%&\hat P^{(2)}_{j+1}  =\small
%\begin{pmatrix}
%   1      &  &  & & & \\
%         & -1  &  & & & \\
%    &   & -1 & & & \\
% &   &  & 1 &  & \\
%     &   &  &  & 1 & \\
%     &   &  & & & -1\\
%     \end{pmatrix}
%\end{split},
%\end{equation}
When applied jointly to two neighboring sites, these operators yield a positive outcome if the link maintains the correct rishon parity and a negative outcome if the link deviates from the correct parity sector. Specifically, $\hat D^{(L)}_n\hat D^{(R)}_{n+1}|\Psi_{\rm phys}\rangle=|\Psi_{\rm phys}\rangle$ and \mbox{$\hat D^{(L)}_n\hat D^{(R)}_{n+1}|\Psi_{\rm unphys}\rangle=-|\Psi_{\rm unphys}\rangle$}, where $|\Psi_{\rm phys}\rangle$ and $|\Psi_{\rm unphys}\rangle$ represent the physical and unphysical states, respectively. These operators establish a truth table that facilitates the post-selection of measurements conforming to the rishon parity.

We propose two different schemes for post-selecting the data, based respectively on destructive and non-destructive measurements. The first exploits the truth table established by the parity operators to determine which states to retain at the end of the simulation. This selection relies solely on the analysis of the qudit populations at each site, rejecting configurations involving adjacent unphysical states--those with just a single rishon on the link.
However, it's important to note that this procedure does not entirely eliminate the possibility that, in previous time steps, the system may have exited the correct parity sector and returned to it afterward. Nevertheless, it serves as a filter and, since it only requires the qudit populations for each time step during the readout, it does not incur additional computational costs.
The second protocol exploits an ancilla qubit to conduct state-preserving measurements.
This approach offers the advantage of continuously monitoring the rishon parity during the time evolution but demands higher resources. The details of this scheme are discussed in App.~\ref{App.link_sym}.
%An alternative protocol based on state preserving measurements, which demand higher resources, is discussed in App.~\ref{App.link_sym}.
%It is noteworthy that for this analysis, only the qudit populations for each time step where readout is performed are necessary, incurring no additional computational cost. 
 %{\color{red} In the previous version, the text was misleading, and the proposed protocol was not clear. I have elaborated  the text to provide more clarity.}

Another quantity that can be exploited to filter the data in post-selection is the total baryon number $\hat{N_b}$, as defined in Sec.~\ref{Sec.ModelB}. This quantity is conserved by the dynamical evolution of the Hamiltonian~\eqref{eq:Heff}. Any population configuration signaling deviations of this quantity from the initial setting of the simulation indicates the occurrence of an error in the evolution, and the corresponding data should be rejected. Also, for this quantity, it is possible to design a non-destructive measurement scheme to monitor possible deviations during the simulation. Unlike the link parity protocol, this approach requires the utilization of an ancilla qutrit instead of a qubit (see App.~\ref{App.link_sym}).

\section{Experimental considerations}\label{Sec.experiment}
In this section, we explore the experimental challenges associated with realizing a quantum digital simulation. Several sources of error can impact the accuracy and effectiveness of the simulation. These include deviations from the Lamb-Dicke regime and the rotating-wave approximation, motional heating, and magnetic field fluctuations~\cite{low2020practical,roos1999quantum,schindler2013quantum}. Another significant hurdle is the implementation of the generalized MS gates via simultaneously driving multiple transitions. Achieving the correct target operation requires the precise calibration of many (linear in the number of transitions) coupled control parameters. Most notably, optical Stark shifts are induced by each laser tone~\cite{haffner2003precision,kirchmair2009deterministic}, which creates an additional layer of complexity. In the following, we discuss some of these experimental challenges, which must be overcome for the successful experimental realization of the proposed quantum digital simulation. 

\subsection{Magnetic field fluctuations}
To evaluate the impact of external magnetic field fluctuations, we assumed infinitely correlated noise in time and space:
throughout each single realization of the dynamics, the magnetic field $\mathcal{B}^z$ perceived by the ion qudits is constant and uniform.
Therefore, the produced Zeeman shifts correspond to
$\hat H_{Z0}= - \mathcal{B}^z \sum_n \hat{\vb{m}}^{z}_n$
with the magnetic dipole operator
\begin{equation}
 \hat{\vb{m}}^z= \frac{1}{2}
 \begin{pmatrix}
   -3 \mu_D &   &   & & & \\
     & -\mu_D &   & & & \\
     &  & -\mu_S & & & \\
     &  &   & \mu_S & & \\
     &  &   & & \mu_D & \\
     &  &   & & & 3 \mu_D \\
\end{pmatrix},
 %{\rm diag}(-3\Delta E_P,-\Delta E_P,-\Delta E_S,\Delta E_S,\Delta E_P,3\Delta E_P)\, , 
\end{equation}
where we consider a ratio $w = \mu_D/\mu_S \approx 0.6$ between the magnetic moments of the D$_{5/2}$ and the S$_{1/2}$ spin-orbitals, compatible with Ref.~\cite{ringbauer2022universal}.
Once we express this Hamiltonian shift in natural units, just as we did for the model Hamiltonian in Eq.~\eqref{eq:Heff}, we can similarly identify a magnetic field realization via the dimensionless parameter
$b = \frac{2 \sqrt{2} a_0 \mu_s}{c \hbar} \mathcal{B}^z$. Then, the rescaled (dimensionless) Hamiltonian
$\hat H_{Z} =  \frac{4 \sqrt{2} a_0}{c \hbar} \hat H_{Z0}$
simply reads
$\hat H_{Z} = -b \sum_{n} \hat{F}_n$ with
\begin{equation}
 \hat F_n = {\rm diag}(-3w,-w,-1,1,w,3w)\, , 
\end{equation}
%shifts of $\pm \Delta E_S$ the energy levels of the states $|3\rangle$ and $|4\rangle$ embedded in the $S$ electronic ground state and of $\pm \Delta E_P$ and $\pm 3\Delta E_P$ respectively the energies of the states $|2\rangle$ and $|5\rangle$
%and $|1\rangle$ and $|6\rangle$ encoded in the $P$ electronic state  $\Delta E_P\approx 0.6 \Delta E_S$.
and we incorporated the magnetic fluctuations Hamiltonian into the Suzuki-Trotter evolution of the Dirac vacuum depicted in Fig.~\ref{fig:circuit}(a). We averaged over 100 realizations of time evolution, while randomly choosing in each realization $r$ the (static, uniform) rescaled magnetic coupling $b_r$ from a uniform (flat) distribution within the interval $b_r \in [-\Delta b,\Delta b]$.
For a comparative analysis with the exact evolution, we computed the state fidelity, $\mathcal{F}(\tilde t)$, at the time $\tilde t$, corresponding to the initial peak in the particle density observed in Fig.~\ref{fig:circuit}(a).
This fidelity is shown in Fig.~\ref{fig:fidelity}(a) as a function of the strength of the magnetic field fluctuations, $\Delta b$, for both first and second-order (see App.~\ref{App.ST2}) Suzuki-Trotter evolutions, considering the same Trotter steps as in Fig.~\ref{fig:circuit}.
%Zeeman-induced ground state energy shift for the two previously considered Rabi frequency-detuning ratios.
The figure demonstrates that the fidelity is not significantly impacted for magnetic fluctuations inducing level shifts up to  $10\%$ of the particle hopping rate, set by $(\eta\Omega)^2/(2|\nu-\delta|)$ in the digital simulation.
%level shifts induced by the magnetic fluctuations up to  $10\%$ of the particle hopping rate in the digital simulation, set by $(\eta\Omega)^2/(2|\nu-\delta|)$,
 %do not significantly impact fidelity. 
For expected hopping rates on the order of $0.1-1$ kHz, this implies a requirement for magnetic field fluctuations to be maintained below the threshold of nT. Achieving this level of precision can be realized with current magnetic shielding techniques~\cite{ruster2016long}. Another strategy that can be employed to reduce magnetic field fluctuations relies on performing a dynamical decoupling scheme~\cite{viola1999dynamical,pelzer2023multi}. In our case, this could be realized by applying two consecutive Suzuki-Trotter steps: the first with the current encoding and the second with a locally dressed basis (see App.~\ref{App.basis_change}) rotated to acquire an opposite magnetic shift. Other schemes to perform dynamical decoupling of qudits, albeit in different ion species, have been recently proposed \cite{farrell2023preparations,zalivako2023continuous}. Nevertheless, the main limitations for our proposal currently rely on intrinsic gate errors and pulse calibration, which we address in the next section.

\subsection{Impact of other errors in the simulation}

\begin{figure}[!t]
    \includegraphics[width=0.49\textwidth]{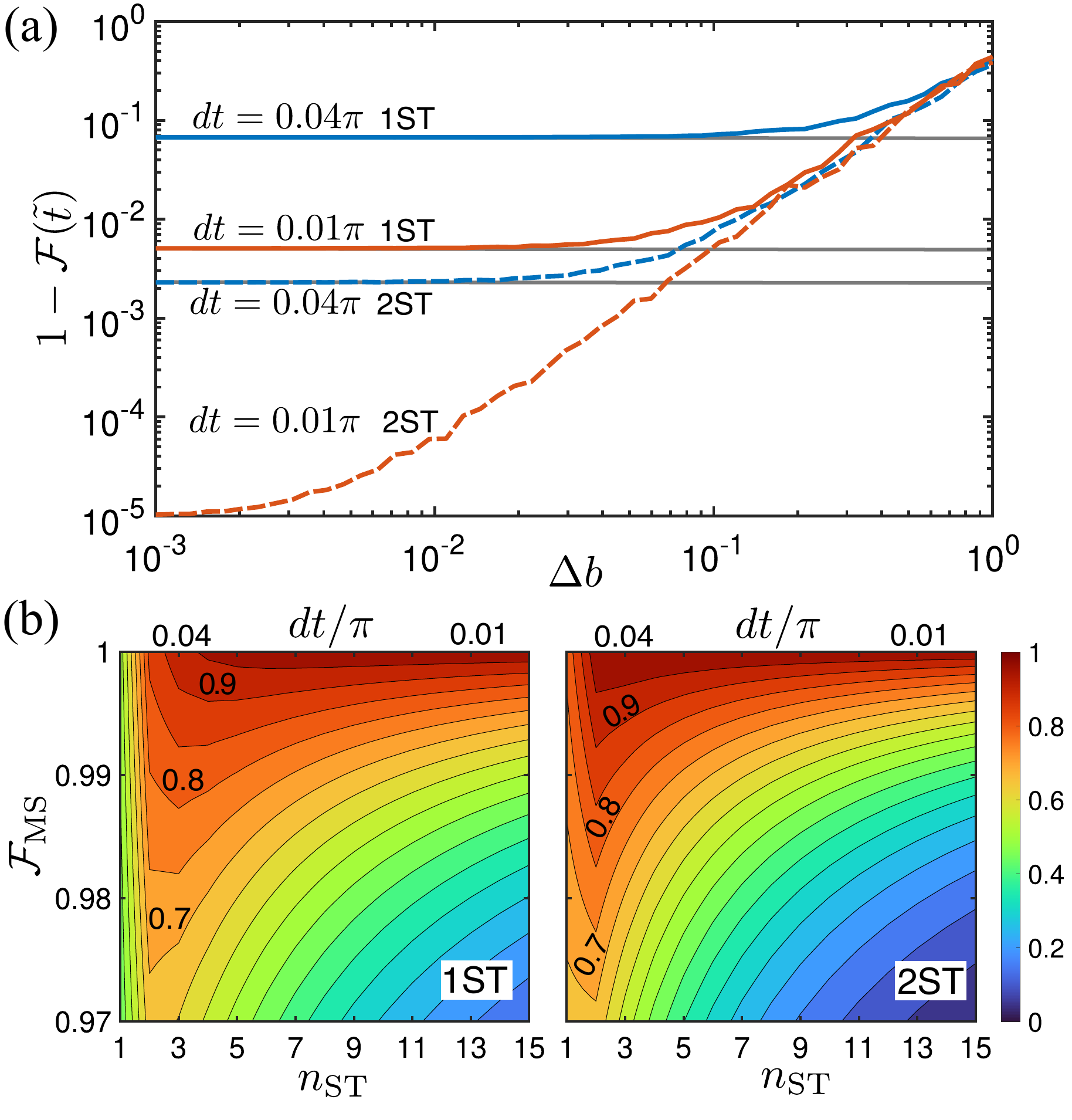}%
    \caption{\textit{Errors estimation}.--(a) State infidelity calculated with respect to the expected state at time $\tilde t=0.377$ and the digitally simulated one as a function of the strength of the magnetic field fluctuations for the same configuration as in Fig.~\ref{fig:circuit}(a). The results were obtained using the first scheme given in Fig.~\ref{fig:circuit}(a) and the continuous and dashed lines represent respectively the results obtained with a first (ST1) and second order (ST2) Suzuki-Trotter evolution. The grey lines indicate the ideal performance obtained in the absence of a magnetic field. We consider the same Suzuki-Trotter time steps as in~\ref{fig:circuit}, which require respectively $n_{\rm ST}=12$ and $n_{\rm ST}=3$ steps to reach $\tilde t$. (b) Estimation of the success probability to reach the peak in the particle density at $\tilde t=0.377$  versus the number of Suzuki-Trotter steps and the fidelity of a single generalized MS gate. The left and right panels are obtained respectively for the first and second-order Suzuki-Trotter evolution. }\label{fig:fidelity}
\end{figure}

 To make an agnostic estimation of the impact of other sources of error in the digital simulation, we assume that a  single generalized two-qudit MS gate occurs with a finite Fidelity $\mathcal{F_{\rm MS}}$. If we neglect errors coming from single qudit operations, which are performed with high precision, the overall performance of the simulation at a given time $t=n_{\rm ST}dt$ can be estimated by the following success probability:
\begin{equation}
    \mathcal{P}(t)=\mathcal{F}(t)[\mathcal{F_{\rm MS}}]^{N_{\rm gates}}\, ,
\end{equation}
where the total number of gates $N_{\rm gates}$ is  defined in  Eq.~\eqref{eq:ngate}.
In Fig.~\ref{fig:fidelity}(b) we plot this success probability as a function of different values of the two-qudit gate fidelity and to the number of Suzuki-Trotter steps necessary to reach, as before, the peak of the particle density creation at time $\tilde t$. In particular, we compare the results obtained with a first and a second-order Suzuki-Trotter evolution, with the latter having $\mathcal{D}=6$ circuit depth (see App.~\ref{App.ST2}). %This estimation shows that, with a two-qudit gate fidelity of the order of $\mathcal{F_{\rm MS}}\gtrsim 0.99$, within the state of the art for qubit MS gates [], a good simulation performance, $\mathcal{P}\sim 90\%$, sufficient for probing the dynamics of the model, can be achieved.
This estimation shows that, with a two-qudit gate fidelity of the order of $\mathcal{F_{\rm MS}}\gtrsim 0.99$ it is possible to obtain a success probability, $\mathcal{P}\sim 90\%$, sufficient for probing the dynamics of the model. Such two-qudit gate fidelities are achievable under the assumption that the errors associated with each transition are independent of each other. With this assumption, the overall error grows linearly with the number of transitions with respect to the standard MS qubit gate infidelity, which, according to the state of the art, can reach values of the order of $0.2\%$~\cite{PhysRevX.13.041052}. However, in the actual implementation, the generalized MS gate performance will depend on the accumulated errors coming from each driven transition, which generically are mutually dependent. The final gate fidelity then will rely on an optimal calibration process, as discussed in the next section.

%The generalized MS gate performance depends on the accumulated errors coming from each driven transition, which generically will be  mutually dependent as discussed in the next section. 
%A rough estimation of the two-qudit gates error can be done assuming that the errors associated to each transition are independent from each other. With this assumption, the error
 %grows linearly with the number of transitions with respect to the standard MS qubit gate infidelity, which, according to the state of the art,
 %can reach values of the order of $0.2\%$ ~\cite{PhysRevX.13.041052}. According this estimation it should be reasonable to achieve a two-qudit gate fidelity of the order of $\mathcal{F_{\rm MS}}\gtrsim 0.99$, which according the results of Fig.~\ref{fig:fidelity}(b) leads to a simulation performance of the order of $\mathcal{P}\sim 90\%$, sufficient for probing the dynamics of the model.

%Considering that state of the art  qubit MS gates reach fidelities of the order of $\sim 99.8\%$~\cite{PhysRevX.13.041052}

\subsection{Calibration challenges}
Using multiple driving fields, even a standard qubit MS gate requires the calibration of four control parameters. In contrast to local gates, these control parameters are non-linearly correlated and can thus not be calibrated independently. As an example, consider a bichromatic light field, symmetrically detuned from an optical transition, as described in Sec.~\ref{Sec:encoding}. Changing the amplitude of one of the two laser tones non-linearly shifts the transition frequency, which in turn changes the effective detuning and thus coupling strength of the other tone. Moreover, due to the multi-level structure of the ion, these Stark shifts are different for each of the optical transitions. For a qubit MS gate, calibration typically involves four control parameters that can be calibrated iteratively or in an automated fashion using Bayesian techniques~\cite{PRXQuantum.3.020350}.

For generalized MS gates, as discussed in Sec~\ref{Sec:encoding}, the situation becomes significantly more difficult. Not only does a change in one parameter affect the entangling dynamics between the two states connected by the addressed transition, but it now also affects the dynamics of states coupled by a simultaneously driven transition. While manual calibration of such a coupled multi-parameter landscape seems very challenging, Bayesian parameter estimation techniques might be extendable to this scenario, given a suitable cost function. Notably, the second approach discussed in Sec.~\ref{Sec:encoding} mediates these challenges, since it requires simultaneous driving only on disjoint transitions. This makes it possible to reduce the coupling between the parameters and pre-calibrate the operations on the two transitions independently to a large extent.

Finally, we observe that a further difficulty for the calibration process comes from the fact that the two-qudit gates presented in Section~\ref{Sec:encoding} require to drive two different sets of transitions on the two ions $n$ and $n+1$. This issue can be solved by applying a rotation of the local dressed basis via unitary transformations, as discussed in Appendix~\ref{App.basis_change}.

\section{Conclusions and outlook}\label{sec.conclusion}
In summary,  we have considered a  Yang-Mills SU(2) 1D lattice gauge theory with dynamically coupled matter truncated at the lowest levels exhibiting non-trivial dynamics and non-abelian features. %where non-trivial dynamics and non-abelian features still play a role.
We introduced a compact rishon representation, embedding local gauge and fermionic degrees of freedom within a dimension-six Hilbert space, which has a natural encoding
 onto a qudit quantum processor based on $^{40}$Ca$^+$ trapped ions. We presented three different schemes relying on entangling gates based respectively on a single, double disjoint and four simultaneously driven transitions per ion.
 %respectively on a single driven transition, double disjoint driven transitions and four simultaneously driven transitions entangling gates. 
 Relevantly, for the latter, we 
 demonstrated that an efficient digital quantum simulation of the model can be accomplished with a notably short circuit depth. 
This result is facilitated by harnessing the computational advantages inherent in qudits and is based on
the efficacy of generalized M{\o}lmer-S{\o}rensen gates. 
 Note that, as also pointed out in other works~\cite{gonzalez2022hardware,zache2023fermion,popov2023variational,methSimulating2DLattice2023},  just the simple encoding of the model into qudits already brings substantial advantages with respect to qubit-based quantum digital simulations, which usually require larger computation resources and circuit depth and the capability of engineering long-range and/or three or more body interactions~\cite{atas20212,PhysRevD.101.074512,PhysRevD.108.094513,farrell2023preparations,liu2023phases,bauer2023new,PhysRevD.102.094501}.

The proposed qudit encoding scheme, based on the local gauge-matter dressed basis, can be generalized to higher representations, symmetries, and dimensions. The most straightforward extensions consist of including the $j=1$ spin shell of the gauge field in the chromoelectric basis, which is achieved by employing a local dressed basis of total dimension $d=10$~\cite{cataldi202321d}. Similarly, an analogous (1+1)D SU(3) Yang-Mills model (truncated  to the smallest nontrivial representations $(1,0)$ and $(0,1)$ for the gauge field), whose phenomenology is closer to QCD such as 3-quark baryons, can be achieved with a local dressed basis of dimension $d=12$~\cite{rigobello2023hadrons}.
 Considering higher spatial dimensions, 
 the non-abelian rishon representation can be applied describe a $j=1/2$ truncated (2+1)D SU(2) model without dynamical matter (pure theory),
 with a local dressed basis of dimension $d=9$~\cite{cataldi202321d}. 
The major challenge in this case would come from the four-body plaquette term associated with the magnetic field. Such interaction could be decomposed into a series of two-qudit gates, as proposed in Ref.~\cite{gonzalez2022hardware}, or exploiting four-body ion interactions~\cite{katz2023demonstration,katz2023programmable}.
All these extensions are within the state-of-the-art development with ion qudits, where full single qudit control of 13-level trapped ion qudits made of  $^{137}\textrm{Ba}^+$ has been recently demonstrated~\cite{low2023control}.
Further extensions could be envisioned by exploiting qudits with larger dimensions, such as those encoded 
 in metastable states of ion isotopes with nuclear spin~\cite{benhelm2008experimental} and
in circular levels of Rydberg atoms~\cite{kruckenhauser2022high,cohen2021quantum}.
%The hardcore gluon approximation can in principle be relaxed while keeping an analogous encoding~\cite{cataldi202321d}, effectively softening the truncation of the gauge fields. 
 %However, such an operation would require the inclusion of additional electronic levels in the programmable atomic quantum processing, potentially achievable with different atomic species~\cite{low2023control}.
 In conclusion, this proposal provides an experimentally feasible and potentially scalable pathway for observing non-abelian lattice gauge phenomena, such as those from high-energy physics or emerging in condensed matter models, in currently available qudit quantum processors.

\section{Acknowledgments}
The authors thank Guido Pagano, Michael Meth,  Jad Halimeh, Giovanni Cataldi and Marco Rigobello for valuable discussions. They acknowledge financial support from:
% EU (QUANTERA + FLAGSHIP + PNRR)
the European Union via QuantERA2017 project QuantHEP,
via QuantERA2021 project T-NiSQ,
via the Quantum Technology Flagship project PASQuanS2,
and via the NextGenerationEU project CN00000013 - Italian Research Center on HPC, Big Data and Quantum Computing (ICSC);
% MUR (PRIN + Dip. Eccellenza)
the Italian Ministry of University and Research (MUR) via PRIN2022-PNRR project TANQU,
and via Progetti Dipartimenti di Eccellenza projects Frontiere Quantistiche (FQ) and Quantum Sensing and Modelling for One-Health (QuaSiModO);
% QCSC
the WCRI-Quantum Computing and Simulation Center (QCSC) of
Padova University.
%G.M. is partially supported by the Italian funding within the ``Budget MUR - Dipartimenti di Eccellenza 2023-2027" (Law 232, 11 December 2016) - Quantum Sensing and Modelling for One-Health (QuaSiModO),
% UNIBA
the University of Bari via grant 2023-UNBACLE-0244025;
% INFN
the Istituto Nazionale di Fisica Nucleare (INFN) via projects NPQCD and Iniziativa Specifica (IS) QUANTUM.
The authors also acknowledge computational resources by Cloud Veneto and the ITensors Library for the MPS calculations~\cite{ITensor,ITensor-r0.3}. The UIBK team acknowledges funding by the European Union under the Horizon Europe Programme--Grant Agreement 101080086--
NeQST, by the European Research Council (ERC, QUDITS, 101080086), and by the Austrian Science Fund (FWF) through the SFB BeyondC (FWF Project No. F7109) and the EU-QUANTERA project TNiSQ (FWF Project No. N-6001). Views and opinions expressed are however those of the author(s) only and do not necessarily reflect those of the European Union or European Climate, Infrastructure and Environment Executive Agency (CINEA). Neither the European Union nor the granting authority can be held responsible for them.
%G.~C. acknowledges that results incorporated in this standard have received funding from  the T-NiSQ consortium agreement financed by QUANTERA 2021.

%\section*{Data availability} The data produced in this manuscript can be provided upon request.

\appendix

\section{Rishon representation}\label{App_A}

Once the hardcore gluon approximation is adopted, the parallel transporter $\hat U^{a b}$ reads, in the chromoelectric 5-dimensional basis of the gauge field space \cite{ZoharBurrelloPRD},
\begin{equation}
    \hat{U}^{a b}=\frac{1}{\sqrt{2}}
    \left(
      \begin{array}{@{\hspace{0.1ex}}c@{\hspace{0.1ex}}|@{\hspace{0.1ex}}c@{\hspace{0.2ex}}c@{\hspace{0.2ex}}c@{\hspace{0.2ex}}c@{\hspace{0.1ex}}}
    0&+\delta_{a\rla}\delta_{b\gla} 
     &-\delta_{a\rla}\delta_{b\rla}  
     &+\delta_{a\gla}\delta_{b\gla}
     &-\delta_{a\gla}\delta_{b\rla} \\
    \hline
    -\delta_{a\gla} \delta_{b\rla}&0&&&\\
    -\delta_{a\gla} \delta_{b\gla}&&0&&\\
    +\delta_{a\rla} \delta_{b\rla}&&&0&\\
    +\delta_{a\rla} \delta_{b\gla}&&&&0\\
  \end{array}
   \right).
    \label{parallel_transport}
\end{equation}
Here we briefly decompose it as a pair of exotic fermion operators, each one acting on a sub-orbital (rishon).
First of all, we need a practical strategy to define exotic fermion operators: A valid fermionic operator has a local action $\tilde{F}$ that inverts a local parity $\hat P = \hat P^{\dagger} = \hat P^{-1}$, so that
$\{\hat P, \tilde{F} \} = 0$.
In analogy to the Jordan-Wigner transformation, the global action of the fermion operator $\hat{F}$ is the string
\begin{equation}
    \hat{F}_n=\dots \hat P_{n-2} \otimes \hat P_{n-1}\otimes \left( \tilde{F} \right)_n \otimes \mathbb{1}_{n+1}\otimes \mathbb{1}_{n+2}\dots ,
    \label{eq:jwstring}
\end{equation}
in contrast to a boson $\hat{B}$ (or spin) global action which reads instead
\begin{equation}
    \hat{B}_n=\dots \mathbb{1}_{n-2} \otimes \mathbb{1}_{n-1}\otimes \left( \tilde{B} \right)_n \otimes \mathbb{1}_{n+1}\otimes \mathbb{1}_{n+2}\dots .
\end{equation}

This prescription ensures that any two fermion operators on different modes anticommute as they should
$\{\hat{F}_n,\hat{F}_{ '\neq n} \}
= \{\hat{F}_n,\hat{\psi}^{(\dagger)}_{n'\neq n} \} = 0$.
In this formalism, it is straightforward to define a Dirac fermion lattice field
\begin{align}
  \hat{\psi}_{\text{Dirac}} &= \qty( 
  \begin{array}{cc}
   0 & 1 \\
   0 & 0
   \end{array})_F&
   \hat P_{\psi} &= \qty( 
  \begin{array}{cc}
   +1 & 0 \\
   0 & -1
   \end{array})  
\end{align}
as well as a Majorana one
\begin{align}
  \hat{c}_{\text{Majo}} &= \qty( 
  \begin{array}{cc}
   0 & 1 \\
   1 & 0
   \end{array})_F&
   \hat P_{c} &= \qty( 
  \begin{array}{cc}
   +1 & 0 \\
   0 & -1
   \end{array})  
\end{align}
where the subscript `$F$' specifies that the operator is meant to be understood as a fermion, i.e.~that his global action has a parity string attached as in Eq.~\eqref{eq:jwstring}.

We can then define a pair of exotic fermion operators for the rishon space, written in the basis
$\{ |0\rangle, |\rla\rangle, |\gla\rangle \}$, namely
\begin{align}
  \hat{\zeta}^{\rla} &= \qty( 
  \begin{array}{c|cc}
   0 & 1 & 0 \\
   \hline
   0 & 0 & 0 \\
   1 & 0 & 0 \\  
   \end{array})_{F}&
  \hat{\zeta}^{\gla} &=
  \qty( 
  \begin{array}{c|cc}
   0 & 0 & 1 \\
   \hline
   -1 & 0 & 0 \\
   0 & 0 & 0 \\  
   \end{array})_{F}
   \label{zeta_definition}
 \end{align}
 with the corresponding local fermion parity
 \begin{equation}
  \hat P_{\zeta} = \qty( 
  \begin{array}{c|cc}
   1 & 0 & 0 \\
   \hline
   0 & -1 & 0 \\
   0 & 0 & -1 \\  
   \end{array}).
   \label{rishon_parity}
\end{equation}
One can then check that the parallel transporter can be decomposed as
\begin{equation}
 \hat{U}^{ab}_{n,n+1} =
 \frac{1}{\sqrt{2}}
 \left( \hat \zeta^a_n \right)_{L}
 \left( \hat \zeta^b_{n+1}\right)_{R}^{\dagger}.
\end{equation}

\begin{table}[t]
 \begin{tabular}{|p{110pt}|p{120pt}|}
  \hline
  \cn $\hat Q_L$ & $\qquad \qquad \qquad \hat Q_R$ \\
  \hline
  \cn $
  \begin{pmatrix}
   0      &  &  & \sqrt{2} & & \\
         & 0  & 1 & & & \\
    &   & 0 & & & 1 \\
 &   &  & 0 & \sqrt{2} & \\
     &   &  &  & 0 & \\
     &   &  & & & 0\\
\end{pmatrix}$
&
  $\quad \;
  -i \begin{pmatrix}
   0      &  & \sqrt{2} &  & & \\
         & 0  & & 1 & & \\
    &   & 0 & & \sqrt{2}  & \\
 &   &  & 0 & & 1 \\
     &   &  &  & 0 & \\
     &   &  & & & 0\\
\end{pmatrix}$
\\ \hline
 \end{tabular}
\caption{ \label{tab:effmatrices3}
 The effective model matrices $\hat{Q}_L$ and $\hat{Q}_R$ written in the gauge-invairant canonical basis of the dressed site.
}
\end{table}
Performing such decomposition greatly simplifies the covariant hopping from the lattice Yang-Mills Hamiltonian, specifically
\begin{equation}\label{eq:hop}
\begin{aligned}
\hat H_{\text{hop}} &= \frac{c\hbar}{2a_0}\sum_n\sum_{a,b=\rla, \gla}\left[-i\hat\psi^{\dagger}_{na}\hat U^{ab}_{n,n+1}\hat\psi_{n+1b}+{\rm H.c.} \right] 
 \\ &=
 \frac{\sqrt{2} c\hbar}{4 a_0}\sum_n\sum_{a,b=\rla, \gla}\left[-i\hat\psi^{\dagger}_{na}
  \left( \hat \zeta^a_n \right)_{L}
   \left( \hat \zeta^b_{n+1} \right)_R^{\dagger}
 \hat\psi_{n+1b}+{\rm H.c.} \right] 
 \\ &=
 \frac{\sqrt{2} c\hbar}{4 a_0}\sum_n
 \left[ 
 \hat{Q}_{Ln}^{\dagger}
 \hat{Q}_{R,n+1}
 + \hat{Q}_{Ln}
 \hat{Q}_{R,n+1}^{\dagger}
 \right]
\end{aligned}
\end{equation}
where the operators
$\hat{Q}_{L,n} = 
\sum_{a=\rla, \gla}
  \left( \hat\zeta^a_n \right)^{\dagger}_{L}
\hat\psi_{na}$ 
and
$\hat{Q}_{R,n} = 
-i \sum_{a=\rla, \gla}
\left( \hat \zeta^a_n \right)^{\dagger}_{R}
\hat\psi_{na}$ 
are explicitly gauge invariant and genuinely local (preserves the fermion parity on the dressed site). On the 6-dimensional logical basis for the dressed site, these matrices read as reported in Table \ref{tab:effmatrices3}. They are not hermitian, however, so they still need to be manipulated for quantum simulation with trapped ions.
Therefore we substitute
\begin{equation}
 \begin{aligned}
  \hat{A}^{(1)} &= \hat{Q}_L + \hat{Q}^{\dagger}_L 
  &
  \hat{B}^{(1)} &= \hat{Q}_R + \hat{Q}^{\dagger}_R
  \\
  \hat{A}^{(2)} &= i (\hat{Q}_L - \hat{Q}^{\dagger}_L ) \quad
  &
  \hat{B}^{(2)} &= i ( \hat{Q}_R - \hat{Q}^{\dagger}_R )
 \end{aligned}
\end{equation}
where we defined four hermitian operators. Now, by noticing that
$\hat{A}^{(1)} \otimes \hat{B}^{(1)} +
\hat{A}^{(2)} \otimes \hat{B}^{(2)}$ is equal to
$
2\hat{Q}^{\dagger}_L \otimes \hat{Q}_R +
2 \hat{Q}_L \otimes \hat{Q}_R^{\dagger}$ we can conclude
that
\begin{equation}\label{eq:hopfin}
\hat H_{\text{hop}} =
 \frac{\sqrt{2} c\hbar}{8 a_0}\sum_n
 \left[ 
 \hat{A}_{n}^{(1)}
 \hat{B}_{n+1}^{(1)}
 +
 \hat{A}_{n}^{(2)}
 \hat{B}_{n+1}^{(2)}
 \right].
\end{equation}
Which is the expression for the nearest-neighbor term which appears in the main text.

Regarding the chromoelectric energy density, we can split the energy contribution half-half to each rishon (the quadratic Casimir eigenvalue must be the same on the two halves due to the link law). Basically
\begin{equation}
\hat H_{\text{elec}} =
 g^2 \frac{c\hbar}{4 a_0}\sum_n
   \left(
   | \hat{\mathbf{R}}_{n,n+1}|^2
   +
   | \hat{\mathbf{L}}_{n,n+1}|^2
   \right),
\end{equation}
but the quadratic Casimir on the rishon space simply reads
 \begin{equation}
  | \hat{\mathbf{L}}|^2 \backslash | \hat{\mathbf{R}}|^2
  = \qty( 
  \begin{array}{c|cc}
   0 & 0 & 0 \\
   \hline
   0 & 3/4 & 0 \\
   0 & 0 & 3/4 \\  
   \end{array}) = \frac{3}{4} \hat{K}.
\end{equation}
Therefore, in conclusion
\begin{equation}
\hat H_{\text{elec}} =
 g^2 \frac{ 3 c\hbar}{16 a_0}\sum_n
   \left(
   \hat{K}_{L,n} + \hat{K}_{R,n}
   \right),
\end{equation}
and the sum $\hat{K}_{L} + \hat{K}_{R}$ on a dressed site is exactly the operator $\hat{C}$.

\begin{figure*}[!t]
    \includegraphics[width=0.95\textwidth]{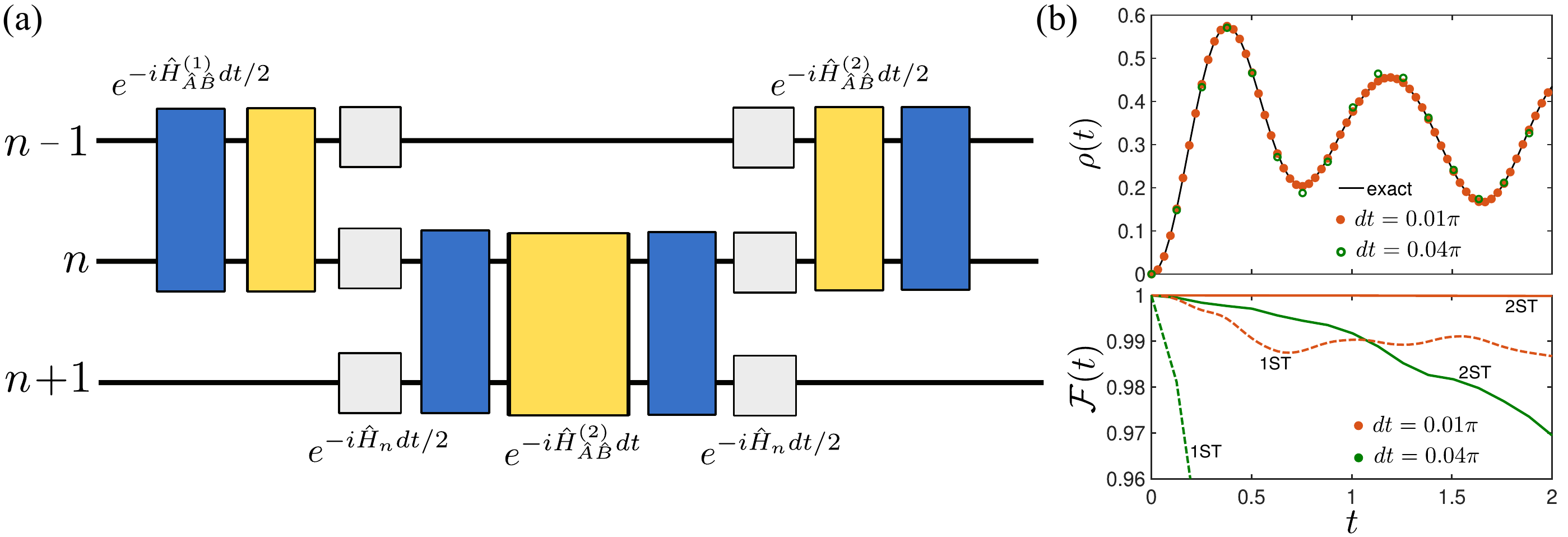}%
    \caption{\textit{Second order Suzuki-Trotter decomposition}.--(a) Unit cell circuit for the second orders Suzuki-Trotter decompostion. The first and last gate can be combined in one of duration $dt$ leading to a circuit depth of $\mathcal{D}=6$. (b) Particle density as function of time  for the Dirac vacuum with $N = 3$ lattice
sites. The continuous line has been obtained with the exact   model while the dots via the second order Suzuki-Trotter  (ST2) evolution of the illustrated circuit using the time steps indicated in the figure. Below is shown the corresponding fidelity between the exact and the digitally simulated state comparing the results of the first (dashed line) and second order (continuous line)  Suzuki-Trotter decomposition. 
For the ST2 we  we included the phononic mode  truncating at the eighth level.% and we set the  Rabi frequency-detuning
%ratios respectively to: $\eta \Omega/|\nu-\delta|=0.1414$ and $\eta \Omega/|\nu-\delta|=0.0708$. 
}\label{fig:ST2}
\end{figure*}

\section{Perturbative model in the strong coupling limit}\label{App.Bar_diff}
In the strong coupling limit, $g\gg 1$, the excitation of the link becomes energetically costly and the field can be adiabatically eliminated. The model's description is then reduced to the bare matter states  $| 1\rangle$ and $| 5\rangle$, which we  rename $|\uparrow\rangle=| 5\rangle$ and $|\downarrow\rangle=| 1\rangle$. A simple hopping process between these two states can be obtained in perturbation theory and leads to the following Heisenberg model in a staggered magnetic field
\begin{equation}
\hat H_{\rm Heis}=-\frac{4}{\bar g^2}\sum_n\hat{\boldsymbol{\sigma}}_n\cdot\hat{\boldsymbol{\sigma}}_{n+1}+m\sum_n(-1)^{n}\hat\sigma^z_n
\end{equation}
where $\bar g^2=g^2+m$ and $\hat{\boldsymbol{\sigma}}=(\hat\sigma^x,\hat\sigma^y,\hat\sigma^z)$. Within this model the Dirac vaccuum correspond to the antiferromagnetic ground state. Assuming to  initially flip a single spin out of this vacuum, corresponding to the excitation of a bare baryon in the original model, a second order hopping process  occur in the limit of $\bar g^2m\gg1$. This process gives rise to the following tight-binding Hamiltonian:
\begin{equation}
\hat H_{\rm eff}=J_{\rm eff}\sum_n\left(\hat\sigma^+_n\hat\sigma^-_{n+1}+\hat\sigma^-_n\hat\sigma^+_{n+1}\right)
\end{equation}
with hopping rate $J_{\rm eff}=16/\bar g^4m$, which describes the diffusion of bare baryons at a speed $v_g=2J_{\rm eff}$. 
%This Hamiltonian well describes the dynamics within the limits considered and predicts the usual dispersion $\omega_k=-2J_{\rm eff}\cos(k)$ and meson group velocity  $v_g=2J_{\rm eff}\sin(k)$.
This result was obtained in a two step perturbation process. A more precise effective description can be obtained performing directly a $4$th order perturbation theory in the limit $g,m\gg J$, considering explicitly the two hopping paths accessible to an initially excited baryon:
\begin{equation}
\begin{split}
&| 5\rangle| 5\rangle| 1\rangle \quad E=0
\qquad\qquad\qquad
| 5\rangle| 5\rangle| 1\rangle \quad E=0\\
&| 5\rangle| 4\rangle| 3\rangle \quad E=2m+2g^2
\qquad\;\;\;
| 5\rangle| 4\rangle| 3\rangle \quad E=2m+2g^2\\
&| 5\rangle| 1\rangle| 5\rangle \quad E=4m
\qquad\qquad\quad\;
| 4\rangle| 6\rangle| 3\rangle \quad E=4g^2
\\
&| 4\rangle| 3\rangle| 5\rangle \quad E=2m+2g^2
\qquad\;\;\;
| 4\rangle| 3\rangle| 5\rangle \quad E=2m+2g^2\\
&| 1\rangle| 5\rangle| 5\rangle \quad E=0
\qquad\qquad\qquad
| 1\rangle| 5\rangle| 5\rangle \quad E=0\\
\end{split},
\end{equation}
where we indicated for each state the bare energy cost.
Performing an adiabatic elimination at the fourth order  we finally  obtain the following effective coupling rate
 \begin{equation}\label{Eq_Jeff}
 J_{\rm eff}=\frac{16}{m\bar g^4}+\frac{8}{g^2 \bar g^4},
 \end{equation}
which reduces to the one previously derived in the limit of $g\gg m$.

 \section{Second order Suzuki-Trotter decomposition}\label{App.ST2}
In the main text we showed how employing a second order Suzuki-Trotter decomposition can be beneficial for the simulation, even if it implies a slightly larger circuit depth.
The unit circuit cells  is presented in Fig.~\ref{fig:ST2}(a). The first and last gate can be merged in one during the evolution thus the the circuit depth is $\mathcal{D}=6$. 
%It  has circuit depth $\mathcal{D}=6$ due to the fact that the first and last gate can be merged in one during the evolution. 
To ensure in this scheme a reduced population of the phonic mode we set  $d \bar t=4\pi /|\nu-\delta|$ for all the applied gates. 
In Fig.~\ref{fig:ST2}(a) we show the performance of this scheme considering again the particle creation out of the Dirac vacuum and the corresponding state Fidelity in time comparing the results obtained with the  first  and second order   Suzuki-Trotter decomposition. The plot shows a clear improvement of the simulation performance due to the reduced Suzuki-Trotter error.

\begin{figure}[!t]
    \includegraphics[width=0.49\textwidth]{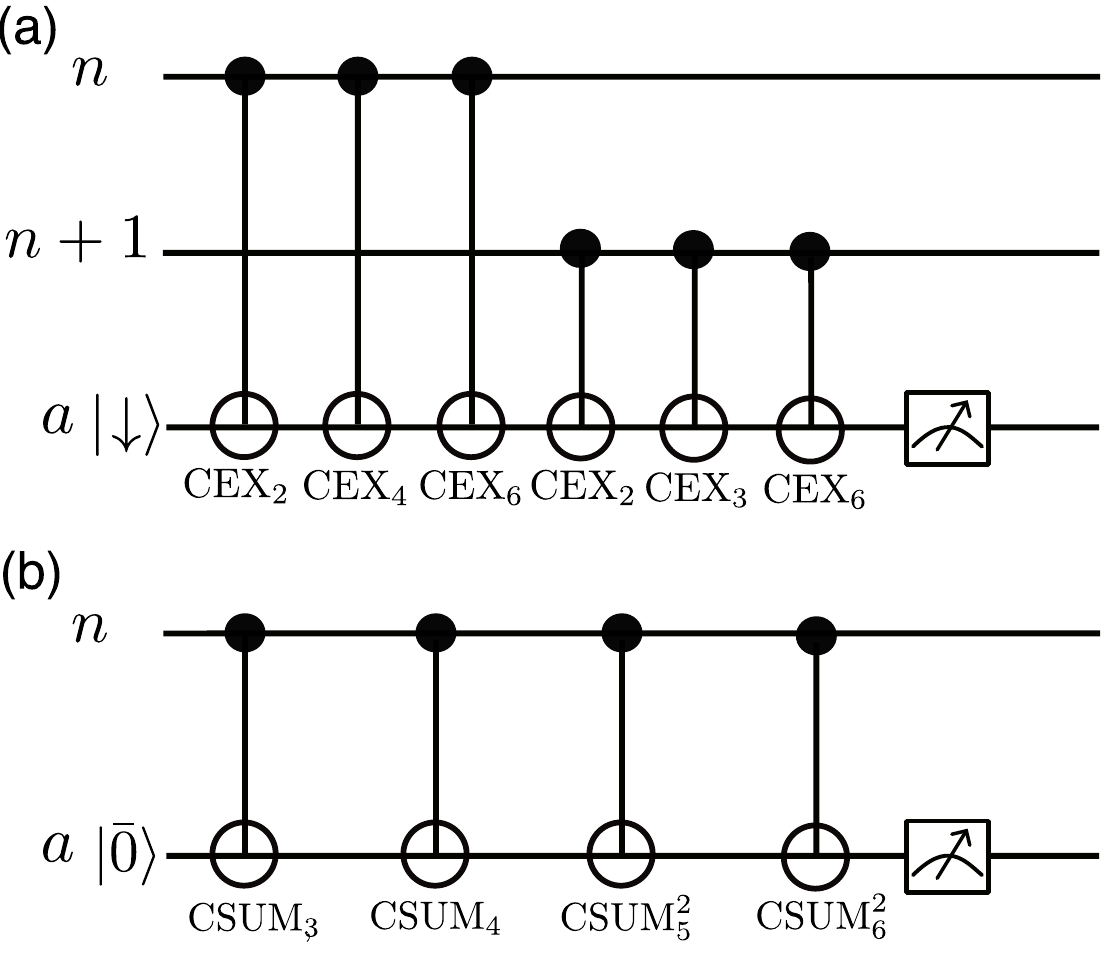}%
    \caption{(a)Parity  and (b) baryon number check circuits to maintain the evolution in the correct parity/baryon sector. In the scheme presented in (a) an ancilla qubit is employed per each pair of qudit while in (b) the same ancilla qutrit is used  for all the chain.} 
   \label{fig:circuit_parity}
\end{figure}

\section{Non-destructive measurement error detection protocol}\label{App.link_sym}

\subsection{Link symmetry}

The link parity preservation protocol presented in the main text to  preserve the link symmetry has the disadvantage of 
relying  on destructive measurements.
As an alternative scheme,  here we propose a strategy  based  on state-preserving measurements. 
This approach offers the advantage of continuously monitoring (and quantum Zeno protecting) rishon parity during the evolution of the system. %It is essential to note that this procedure, in contrast to a straightforward post-selection method, incurs an additional computational cost.
The core concept involves the utilization of an ancilla qubit per qudit pair, which may also be represented by an unused qudit level, to conduct state-preserving measurements. Focusing on two adjacent sites, denoted as $n$ and $n+1$, let us consider a scenario where, during the system's evolution, the state tends to leak out of the parity sector due to experimental errors.
 The generic form of the state of the two sites then reads:
\begin{equation}
|\psi\rangle=\sum_r a_r|\psi_{\rm phys}\rangle_r+\sum_r b_r|\psi_{\rm unphys}\rangle_r
\end{equation}
where $|\psi_{\rm phys}\rangle_r$ and $|\psi_{\rm unphys}\rangle_r$ represent states living in the even and odd symmetry sector. Here for simplicity we are assuming a coherent superposition between the two but an analogue argument works for a generic statistical mixture.
To non-destructively measure parity, we employ a hybrid version of the ${\rm CEX}_c$ gate, already implemented in the trapped ion qudit quantum processor~\cite{ringbauer2022universal}
\begin{equation}
    {\rm CEX}_c=\left\{\begin{array}{cc}
       |s,\uparrow\rangle \leftrightarrow  |s,\downarrow\rangle  & {\rm if}  \;\;s= c \\
      |s,k\rangle \rightarrow  |s,k\rangle  &  {\rm if}  \;\;s\neq c
    \end{array}\right.,
\end{equation}
where $|s,k\rangle$ represents the product state between the qudit and the qubit.
This gate flips the ancilla qubit's state  if the qudit is in the state $|c\rangle$ and retains the qubit's state if the qudit is in any other state. We initialize the ancilla in the state $|\downarrow\rangle_a$ and apply a sequence of ${\rm CEX}_c$ gates on each qudit. The choice of the $|c\rangle$ qudit' s state corresponds to the negative values of the two parity operators given in Table~\eqref{tab:effmatrices2} associated to  each qudit pair,  as illustrated in Fig.~\ref{fig:circuit_parity}(a). A similar sequence is then applied separately at first and last qudits of the chain to ensure that the  parity remains fixed at the boundaries.
 This process results in the following state for two adjacent sites:
\begin{equation}
|\psi\rangle=\sum_r a_r|\psi_{\rm phys}\rangle_r|\downarrow\rangle_a+\sum_r b_r|\psi_{\rm unphys}\rangle_r|\uparrow\rangle_a\,.
\end{equation}

With this procedure, we can monitor the state's evolution  at each time step. In particular, if the ancilla qubit remains in its initial state, we can be confident that the evolution has occurred within the correct symmetry sector. However, if the ancilla qubit flips, indicating a deviation, it suggests a leakage into the odd symmetry sector.

Note that, if we neglect the possibility of having more than one qudit error in the chain within a given time step, this non-destructive measurement of the link parity errors is capable to detect all possible single qudit state flips besides the ones maintaining the same link parity, i.e. $|1\rangle \leftrightarrow |5\rangle$ and 
 $|2\rangle \leftrightarrow |6\rangle$. To detect also this kind of errors in the next section we propose a non-destructive scheme to monitor deviations in the baryon number of the chain.

%also signals deviation in the baryon number $\hat{N_b} = \frac{1}{2} \sum_n (\hat{M}_n - 1)$ from integer to fractional. This is associated with transitions of the kind $\{|1\rangle, |2\rangle, |5\rangle, |6\rangle\} \leftrightarrow \{|3\rangle, |4\rangle\}$. To also exclude qudit flips of the kind $\{|1\rangle, |2\rangle\} \leftrightarrow \{|5\rangle, |6\rangle\}$, in the next section, we illustrate a protocol to monitor deviations in the baryon number of the chain.}
%In response to this deviation, we may applying an $X$ qudit gate defined as:
%\begin{equation}
%X=\sum_j|j+1({\rm mod} \; 6)\rangle\langle j|
%\end{equation}
% on the first qudit. While this operation doesn't correct errors, it effectively project the unphysical state back into the correct symmetry sector. Although not error-correcting per se, this action can assist the system in maintaining the intended dynamics.

 \subsection{Baryon number}
Let us assume, in the following, that our qudit quantum processor is affected only by single qudit errors in each time step. Such errors can induce a modification of the total baryon number $\hat{N_b} = \frac{1}{2} \sum_n (\hat{M}_n - 1)$, which for a single qudit flip can vary by an half integer or by an integer, i.e. $\Delta\hat{N_b} = \pm\frac{1}{2}, \pm 1$. Deviation in this quantity can be monitored during the evolution by using an ancilla qutrit for the entire chain spanned by the basis set $|t\rangle_a = {|\bar{0}\rangle_a, |\bar{1}\rangle_a, |\bar{2}\rangle_a}$, with the two upper states counting for half integer and integer deviations, respectively. In this case, we employ the ${\rm CSUM}_c$ gate, which is available in the trapped ion qudit quantum processor~\cite{ringbauer2022universal}:
\begin{equation}
    {\rm CSUM}_{c}=\left\{\begin{array}{cc}
       \;\; \;\;\,|s,t\rangle \leftrightarrow  |s,t\oplus 1\rangle  & {\rm if}  \;\;s= c \\
      |s,k\rangle \rightarrow  |s,k\rangle  &  {\rm if}  \;\;s\neq c
    \end{array}\right.
\end{equation}
where $\oplus$ denotes addition modulo $3$ and where $|s,k\rangle$ represent the product state between the qudit and the qutrit. We initialize the qutrit ancilla in the state $|\bar{0}\rangle_a$, and then apply to each qudit the sequence of gates ${\rm CSUM}_{3}$, ${\rm CSUM}_{4}$, ${\rm CSUM}_{5}^2$, ${\rm CSUM}_{6}^2$ shown in Fig.~\ref{fig:circuit_parity}(b). Here, the gates ${\rm CSUM}_c^2$ perform the operation $|c, t\rangle \leftrightarrow |c, t\oplus 2\rangle$ if the qudit is in state $c$. With this scheme, if the ancilla qutrit at the end of the process takes a value different from $|\bar{0}\rangle$, a single qudit flip occurred. Importantly, the qudit flips not detectable from the link parity symmetry, i.e., $|1\rangle \leftrightarrow |5\rangle$ and $|2\rangle \leftrightarrow |6\rangle$, are signaled by a qutrit readout $|t\rangle_a = |\bar{2}\rangle$. With this procedure, combined with the link-symmetry non-destructive measurement, all possible single qudit symmetry-violations can be detected at each time step.

\section{Local dressed basis rotation}\label{App.basis_change}
The two-qudit gates discussed in Sec.~\ref{Sec:encoding}
require to drive two different set of transitions on the two ion's qudit.
This is not a fundamental problem per se but it can  affect the complexity of the pulse calibration process.
A solution to this issue consists in performing an unitary transformation of the local dressed basis on just the even lattice sites using the following unitary operator:
\begin{equation}
 \hat V= 
 \begin{pmatrix}
    i &   &   & & & \\
     & i &   & & & \\
     &  & 0 & 1 & & \\
     &  & 1  & 0 & & \\
     &  &   & & -i & \\
     &  &   & & &  -i \\
\end{pmatrix}.
\end{equation}
This matrix transforms  the operators of the model~\eqref{eq:Heff} according to
\begin{equation}
\begin{split}
 &\hat V  \hat A^{(k)} \hat V ^{\dagger}=-\hat B^{(k)}\\
 &\hat V  \hat B^{(k)} \hat  V^{\dagger}=\hat A^{(k)}\\
 & \hat V \hat M \hat V ^{\dagger}=\hat M\\
 &   \hat V \hat C \hat V ^{\dagger}=\hat C
    \end{split}
\end{equation}
with $k=1,2$.
The Hamiltonian given in Eq.~\eqref{eq:Heff} than reads
\begin{equation}
\begin{split}
\hat H= &\sum_{n\in \rm odd}\left[\hat A^{(1)}_n\hat A^{(1)}_{n+1}+\hat A^{(2)}_n\hat A^{(2)}_{n+1}\right]\\
-&\sum_{n\in \rm even}\left[\hat B^{(1)}_n\hat B^{(1)}_{n+1}+\hat B^{(2)}_n\hat B^{(2)}_{n+1}\right]\\
+&m\sum_n(-1)^{n}\hat M_n+g^2\sum_n\hat C_n \,. \\ \\
\end{split}
\end{equation}
In this way, as required, the two-qudit gates necessary to implement the hopping terms always involve the same operator acting on the pair of qudit.
This change of local basis should be applied before and after the sequence of four two-qudit gates presented in Fig.~\ref{fig:circuit} of the main text. 
To conclude the protocol, after having returned to the original basis, the single qudit compensation operations (see Eq.~\ref{eq:corr}) can be applied as in the main text.

For the disjoint double-transition scheme is also possible to apply  local transformations  in order two have two-qudit gates involving the same operator on both qudit. Compared to the full scheme,  the same basis transformation can not be applied for all the gates but  different single qudit rotations should be applied on each site before and after  each two-qudit gate. In particular, it is possible to reduce all the two qudit gates to be always of the same form by using the unitary transformations
\begin{equation}
 \hat V_{\hat\alpha^{(k)}_{q}\hat \alpha^{(k')}_{q'}}  \hat  \alpha^{(k)}_q \hat V_{\hat\alpha^{(k)}_q\hat\alpha^{(k')}_{q'}}^{\dagger}= \hat\alpha^{(k')}_{q'}
 \end{equation}
\begin{equation}
 \hat V_{\hat\beta^{(k)}_{q}\hat\alpha^{(k')}_{q'}}  \hat \beta^{(k)}_q \hat V_{\hat\beta^{(k)}_q\hat\alpha^{(k')}_{q'}}^{\dagger}= \hat \alpha^{(k')}_{q'},
 \end{equation}
with $k,q,k',q'=1,2$.
Again, the single qudit compensation matrices could be applied similar as in the main text at the end of the series of gates when the local basis is transformed back to the original one.

\newpage
%\vspace{15em} 
%Finally note that an alternative procedure to reduce the number of basis transformations could be performed by applying the basis change  only to the even sites. In this case the two qudit gates will not reduced always to the same operators but will be generically of the form $\alpha^{(k)}_{q}\alpha^{(k)}_q$ or $\beta^{(k)}_{q}\beta^{(k)}_q$.

\bibliography{refs}

\end{document}